# Ball Lightning as a Hot, Highly Charged, Sphere of Air

Chris Allen Broka
(chris.broka@gmail.com)

**Abstract.**

An exceptionally simple model of ball lightning is proposed that describes it as a highly charged sphere of hot, conductive, air surrounded by colder air. This conductive sphere possesses a net excess of charge. This charge will create a corona discharge that will heat the surrounding air thus maintaining the temperature of the ball itself. Numerical simulations are presented which give results that would appear to agree with what many witnesses have reported. It is argued that such spheres are likely positively charged.

## 1. Introduction.

Ball lightning is a peculiar atmospheric phenomenon usually, but not always, associated with lightning strikes (Uman 1969; Rakov et. al. 2003). Balls of light are observed to move about, being smaller than a golf ball or even several meters in diameter. Their color is often red, orange, or white. They may last anywhere from several seconds to several minutes but 5 sec seems typical (Rakov et. al. 2003). Their demise is often violent as they frequently explode (either spontaneously or as a result of contact with objects). Many theories have been put forward to explain these strange observations. Some are chemical in nature; burning silicon has been suggested (Abrahamson et. al. 2000). Cen et. al. (2014) recently captured what may well be the spectrum of a ball lightning produced by a cloud-to-ground strike. Lines for silicon, iron, and calcium were observed in addition to lines for neutral atomic nitrogen and oxygen. These latter are not surprising and suggest that the phenomenon is significantly cooler than the parent lightning channel, in the spectrum of which were observed lines for ionized nitrogen and oxygen. The other elements must surely have come from the soil. But it is hard to say whether they played any active role in the ball lightning process. Kapitza (1955) proposed a a model in which microwave radiation is trapped within a spherical cavity. Other authors have extended this idea (Wu 2016). Ranada (2000) has pictured it as a force-free, topologically entangled, magnetic knot (also see (Tsui 2003)). Different suggestions include black holes (Rabinowitz 2002) and Rydberg matter (Manykin et. al. 2006). Morrow (2018) envisions a sphere of positive ions that attracts burning, negatively charged, particles into it. These account for the ball's luminosity.

The present model concentrates on the thermodynamics of a sphere of very hot air at 1 atm. This problem was examined by Lowke et. al. (1969) who found that such spheres would probably not support their temperature or luminosity long enough to be viable candidates for ball lightning. Tesla (1899) had speculated that a small sphere of hot, diaphanous, air might conduct charge from sky to ground, thus maintaining its temperature. Uman et. al. examined a similar idea (1966).

There is considerable evidence to suggest that these lightning balls may be charged (Charman 1972). They are sometimes observed to move towards or along conducting surfaces. Very significantly, they have been described as giving off filamentous or corona discharges as well as hissing sounds. Ozone and acrid smells associated with such discharges are also reported. Several people, for instance Georg Richmann (Clarke 1983), seem to have been electrocuted by the phenomenon. We will take this evidence seriously and propose a model of ball lightning that pictures it as a sphere of very hot, conductive, air bearing an intense charge. The energy generated by the dissipation of this charge will serve to keep the inside of the ball hot.

## 2. A Model of Ball Lightning.

When a lightning strike comes down it will certainly create quite a bit of highly heated and partially ionized air owing to the tremendous energies involved. Suppose that the hot and conductive mass of air produced by this process manifests an imbalance in its contained charge. (We will discuss further how such a situation might come about below). It is not clear, right now, whether the excess charge should be regarded as positive or negative. We imagine this plasma would start off with a temperature on the order of at least ~ 4, 000 $K$. (Note that we are not suggesting that the ball is, literally, formed out of the lightning channel itself. It is, rather, a consequence of the channel's striking down.) Ordinarily, such a mass of heated air would be expected to cool in a very short time (Lowke et. al. 1969). We propose that, because it is highly charged, it remains heated under the influence of its own corona discharge. We will examine and numerically simulate this situation below. While no very detailed mechanism can be provided, at this time, to explain the formation of such a region of hot, highly charged, air it is worth pointing out that a lightning strike generates both charge and highly heated air in great abundance; these are the only raw materials necessary for our model to work. It requires no external source of power. The excess charges are, initially, trapped by the interface between the very hot air and the cold, non-conductive, air outside the ball. We speculate that some of these charges will escape, or be neutralized, in a process somewhat reminiscent of thermionic emission. We are not sure yet if these excess charges are positive ions or electrons. But we assume them to be distributed uniformly throughout the ball. It might be objected that these excess charge carriers would generate a significant electric field and immediately be expelled from the ball by their own electrostatic force. But the air inside the ball is considerably ionized owing to its high temperature and there may be other ions present as well. The Debye length is very small. The excess charge carriers will feel little force from their companions causing them to move outward. This matter will be examined in greater detail presently.

Now an obvious question comes to mind – why is ball lightning round? Suppose that our mass of highly charged air starts out in some irregular shape. Suppose it has a sharp spike or protrusion in some area. Around such a protrusion the electric field will be extraordinarily high and we expect the rate of charge emission there will surely be greater. The surface charge density will drop in this region quite rapidly. Also, it will lose heat more quickly in this area. Likewise, if there were to be a more concave or flat area on the mass, the electric field there would be weaker. The surface charge density would diminish more slowly in that region. In this way the mass would tend to work itself into a situation where its surface charge density was everywhere the same. This, in turn, requires that it be *spherical* in shape. But it would not be very stable. Any serious turbulence in the atmosphere would tear it apart. And, if it came into contact with an object, it would rapidly dissipate. Ball lightning would be favored in areas of calm wind. It would stand a better chance of lasting if it formed rather far away from any objects it might encounter. Such a ball – having almost no density – would be quite buoyant. But it would also, as a charged object, be influenced by local electric fields which might draw it downwards (*vide infra*). Only under those happy circumstances where everything works out right do we see ball lightning. In most cases it probably floats away or crashes into the ground unnoticed.

We assume that the pressure here is everywhere $P_0$ = 1 atm (101, 325 N/$m^2$). $\eta$ (the number density) = $P_0$/k T always. We assume that $\beta$ k T is the average energy of any particular particle ($\beta = \frac{1}{\gamma-1}$). The initial conditions inside the ball, at t = 0, are taken to be those of air at 1 atm pressure and 4, 000 $K$. It may well have been hotter to begin with. But above about 8, $000^o$ the ball will cool very quickly due to its emission of light. And even much above 6, $000^o$ it would glow too brightly to be a realistic ball lightning. At 4, $000^o$ it would consist, primarily, of partially ionized nitrogen and oxygen molecules. Outside will be an area of corona discharge.

Let us suppose that a lightning strike has come down and deposited such a hot sphere with a radius $R_0$ and an initial charge $Q_0$. Outside this charged sphere will be an electric field – perhaps very strong. This will produce some ionization in the air around it. If the sphere is negatively charged electrons will leak away and escape from it. If it is positively charged electrons will flow to it from infinity. This will produce further electrons and positively charged ions outside the ball. Some of these will eventually fall back upon the sphere, thus reducing its charge. Some electrons will neutralize positive ions. There will also be chemical reactions taking place between the various gas molecules and ions in this very energetic region. We have no reliable way of estimating just how fast these processes will occur. But they all seem more-or-less related to the electric field strength at the surface of the ball. So let us guess that $\partial_t Q(t) = -\alpha Q(t)$ where $\alpha$ is some unknown constant and $Q(t)$ is the charge of the ball. $\alpha$ measures the rate of corona discharge as a function of the ball's net charge. Thus we will assume $Q(t) = Q_0 e^{-\alpha t}$.

Now we imagine that each unit of charge generated as a result of this complicated process makes its way from the ball's surface to r = ∞ , or vice-versa, at a sort-of constant drift velocity and that it does so very quickly in relation to the ball's lifetime; the energy it gains from the ball's electrical potential mostly goes into heating the gas it passes through. One can simply picture surrounding the ball with shells of radii (r, r+dr) and asking how much energy the charge gives out in moving through each. One must also consider how many particles there are in each shell that will have to share this energy. Considering $\beta$, one can easily find the local rate of temperature increase. This will be greatest near the ball's surface and much less significant far away. But the local temperature will also change at a rate $\nabla \cdot (D(T(r, t)) \nabla T(r, t))$ in consequence of thermal diffusion where we write the thermal diffusivity as a function of temperature. We find:

1) $$\partial_t T(r, t) = \frac{-(\partial_t Q(t))\, Q(t)}{16 \pi^2 \epsilon_0 r^4 \beta k \eta} + \nabla \cdot (D(T(r, t)) \nabla T(r, t))$$

We suppose that $\partial_t Q(t) \approx -\alpha Q(t)$.

2) $$\partial_t T(r, t) = \frac{\alpha Q_0^2 e^{-2 \alpha t}}{16 \pi^2 \epsilon_0 r^4 \beta P_0} T(r, t) + \nabla \cdot (D(T(r, t)) \nabla T(r, t)).$$

We cannot neglect cooling due to radiative losses. We designate this term $W(T(r, t))$ (W/$m^3$). Therefore:

3) $$\partial_t T(r, t) = U(r - R_0) \frac{\alpha Q_0^2 e^{-2 \alpha t}}{16 \pi^2 \epsilon_0 r^4 \beta P_0} T(r, t) + \nabla \cdot (D(T(r, t)) \nabla T(r, t)) - \frac{W(T(r, t))}{P_0 \beta} T(r, t)$$ where $U(r - R_0)$ represents the unit step function. Heat is only generated outside the ball.

We do not care, at this stage, whether the excess charge is positive or negative since the net energetics are the same in both cases.

## 3. Numerical Simulations.

In order to see whether Equation 3) describes any interesting physics it must be solved numerically. In order to do this we need expressions for $D(T(r, t))$ and $W(T(r, t))$ that are both reasonably accurate and simple for a computer to work with. $D(T(r, t))$ can be estimated by fitting the thermal conductivity ($\kappa$) of very hot air from (Yos 1963 and Engineeringtoolbox.com) to a curve and recalling that $D(T(r, t)) = \kappa/(\rho\, C_P)$.

In this new version of the original paper we introduce an improved method for obtaining the numerical solutions. The solution of this non-linear PDE is not straightforward. It is difficult, using *Mathematica*'s NDSolve function, to capture the expansion of the ball. This is a crucially important matter since it directly influences the rate of the ball's heating. We require a more effective method. This method is adapted from (Broka 2020).

We will construct a table, *K*, that represents the initial state of the ball:

```
K = {{0.01, 4000}, {0.02, 4000}, {0.03, 4000}, {0.04, 4000}, {0.05, 4000}, {0.06, 4000}, {0.07, 4000},
    {0.08, 4000}, {0.09, 4000}, {0.1, 4000}, {0.11, 300}, {0.12, 300}, {0.13, 300},
    {0.14, 300}, {0.15, 300}, {0.16, 300}, {0.17, 300}, {0.18, 300}, {0.19, 300}, {0.2, 300},
    {0.21, 300}, {0.22, 300}, {0.23, 300}, {0.24, 300}, {0.25, 300}, {0.26, 300}, {0.27, 300},
    {0.28, 300}, {0.29, 300}, {0.3, 300}, {0.31, 300}, {0.32, 300}, {0.33, 300}, {0.34, 300},
    {0.35, 300}, {0.36, 300}, {0.37, 300}, {0.38, 300}, {0.39, 300}, {0.4, 300}, {0, 0}}.
```

Here the first term in the elements of the set designates the radial distance from the center of the ball and the second its temperature. We imagine that the ball begins at 4000 K – not too unbelievable – and that the air outside is at 300 K. Essentially, we are treating the ball as a sort of "onion." We assume that nothing ever crosses from one onion-layer to another. Any air molecule or free electron that starts in a given shell stays there forever. Heat is only given to the outermost 30 shells. The 41st element of the set is a 'dummy element' that is only used to index time throughout the simulation. We guess that the initial radius is 10 cm.

We first act on *K* with a 'Source Operator' which adds the proper amount of temperature to each layer of the onion and subtracts the loss due to continuum radiation (usually insignificant). After this a 'Diffusion Operator' acts upon it doing two things. Firstly, it redistributes the temperatures according to ordinary thermal diffusion. Then it adds one time increment to the 'dummy' 31st set. Lastly an operator rearranges the coordinates of the onion-layers so that expansion is accounted for and the pressure restored to $P_0$ everywhere – if, say, the temperature inside a particular layer happened to increase, the separation between that layer's walls would have to increase as well. The expansion/contraction operation is performed adiabatically. These operators execute over a time step small enough that the solution is stable and accurate. We combine these three sequential processes into one 'Evolution Operator' which is iterated upon *K* as many times as necessary to cover the time interval of interest. (The details are laid out in the Supplementary Material.)

We have no idea what value to select for $\alpha$ but we do know that it is related to the ball's lifetime. Suppose our ball starts with a radius of .1 m (which is not atypical of many reported cases). Such a ball might be expected to last on the order of 10 sec. So let us choose $\alpha = .1\ \text{sec}^{-1}$ just to see if anything reasonable happens. Guessing at values for $Q_0$, we find the most interesting results occurring when $Q_0$ is about .0015 C. Much below this there is simply not enough charge to heat the ball sufficiently to prolong its life – the ball quickly cools and dies (fig. 1). It produces little visible light. For $Q_0 = .0015$ the ball quickly cools but then remains close to $3000^o$, and easily visible, for at least 20 sec which brings it well within the range reported by witnesses. Below is plotted its temperature profile as a function of time (fig. 2). We plot the decay of the ball's core temperature with and without charge (fig. 3). (See Supplementary Material *Mathematica.*) Its maximum luminosity is about 1.2 W. This would correspond to the visible light output from a 60 W incandescent light bulb, assuming these are about 2% efficient (Armaroli et. al. 2011). This is not unlike what many witnesses report. Since a lightning strike is believed to usually bring down about 10 C of charge (Uman 1969), a figure of $Q_0 = .0015$ C is hardly unreasonable. If the initial charge had to be 1000 C we would wonder where that much charge came from.

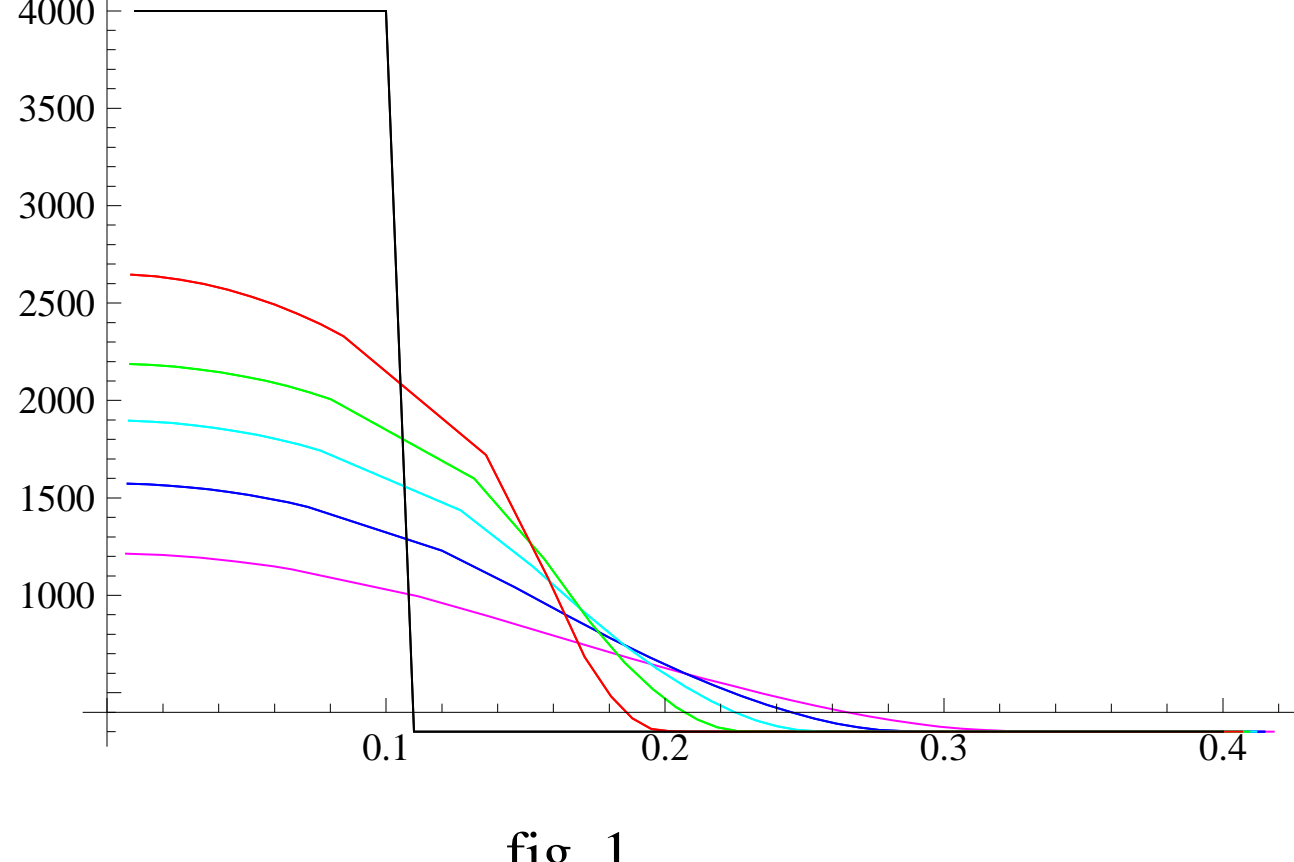

fig. 1

The calculated temperature profile ($K$) of an $R_0 = .1$ m, $Q_0 = 0$ C ball as a function of radius (m) plotted for $t = 0$ (black), 2 (red), 5 (green), 10 (blue-green), 20 (blue), 40 (purple) sec.

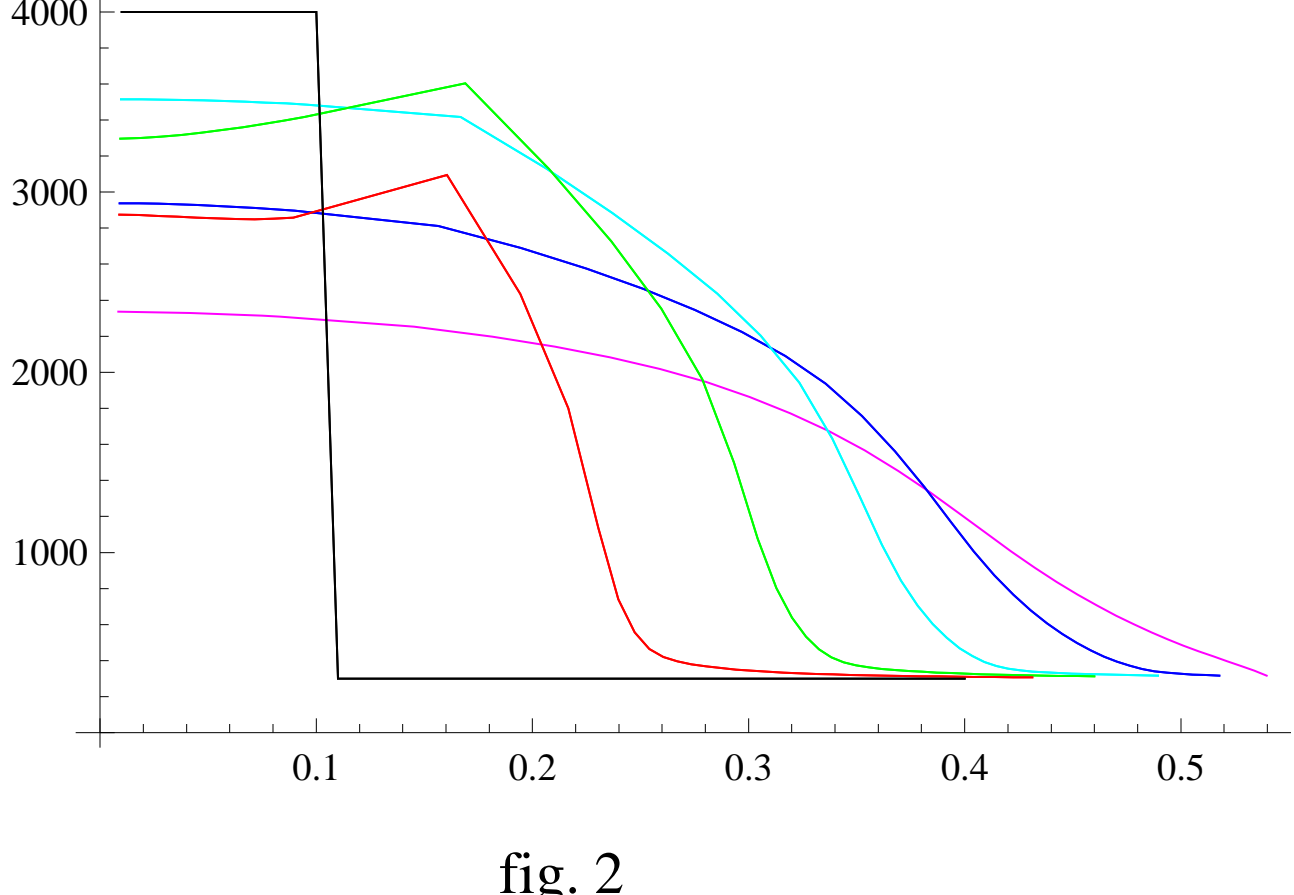

fig. 2

The calculated temperature profile ($K$) of an $R_0 = .1$ m, $Q_0 = .0015$ C, $\alpha = .1$, ball as a function of radius (m) plotted for $t = 0$ (black), 2 (red), 5 (green), 10 (blue-green), 20 (blue), 40 (purple) sec.

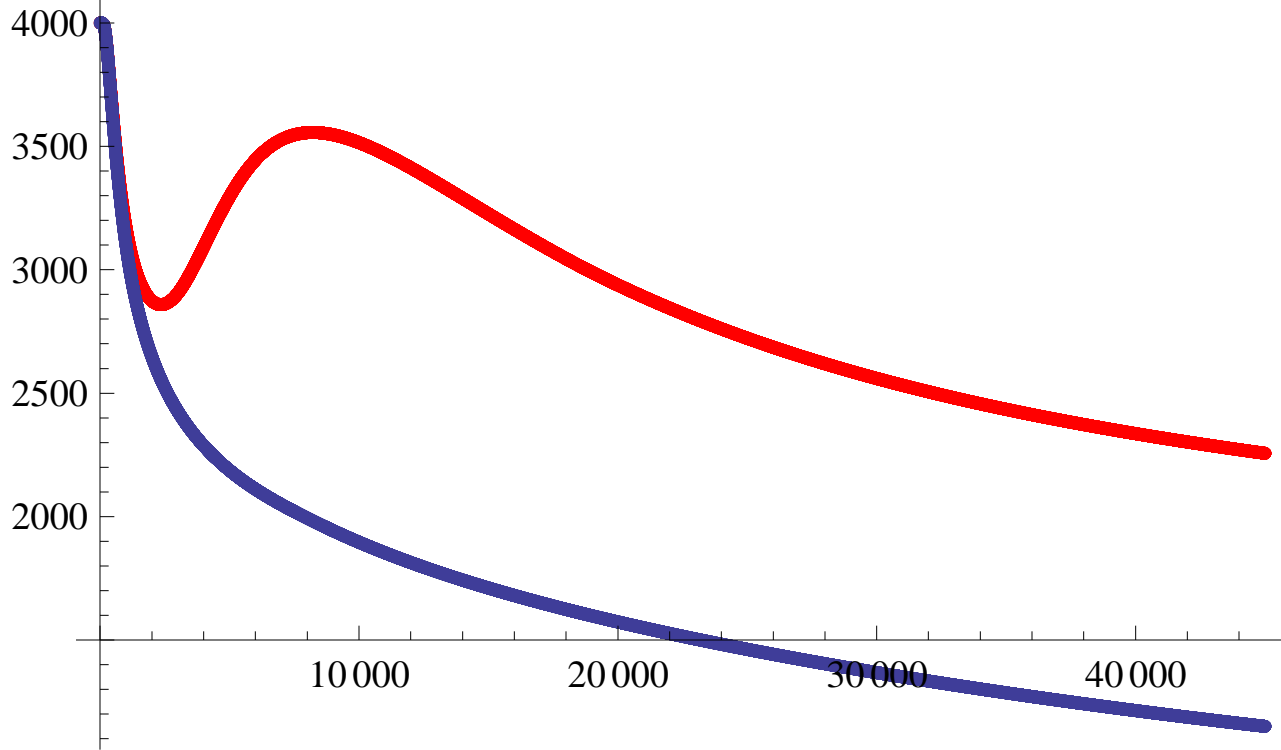

fig. 3

The calculated core temperature ($K$) of an $R_0$ = . 1 m, $Q_0$ = .0015 C (red) and $Q_0$ = 0 (blue), ball as a function of time (msec).

Below (fig. 4) we plot the luminosity profile of this ball as a function of time. As a witness might observe it, it would have no well-defined or hard-edged boundary. Most of the light would come from inside the ball, however. This is generally consistent with reports (Charman 1972).

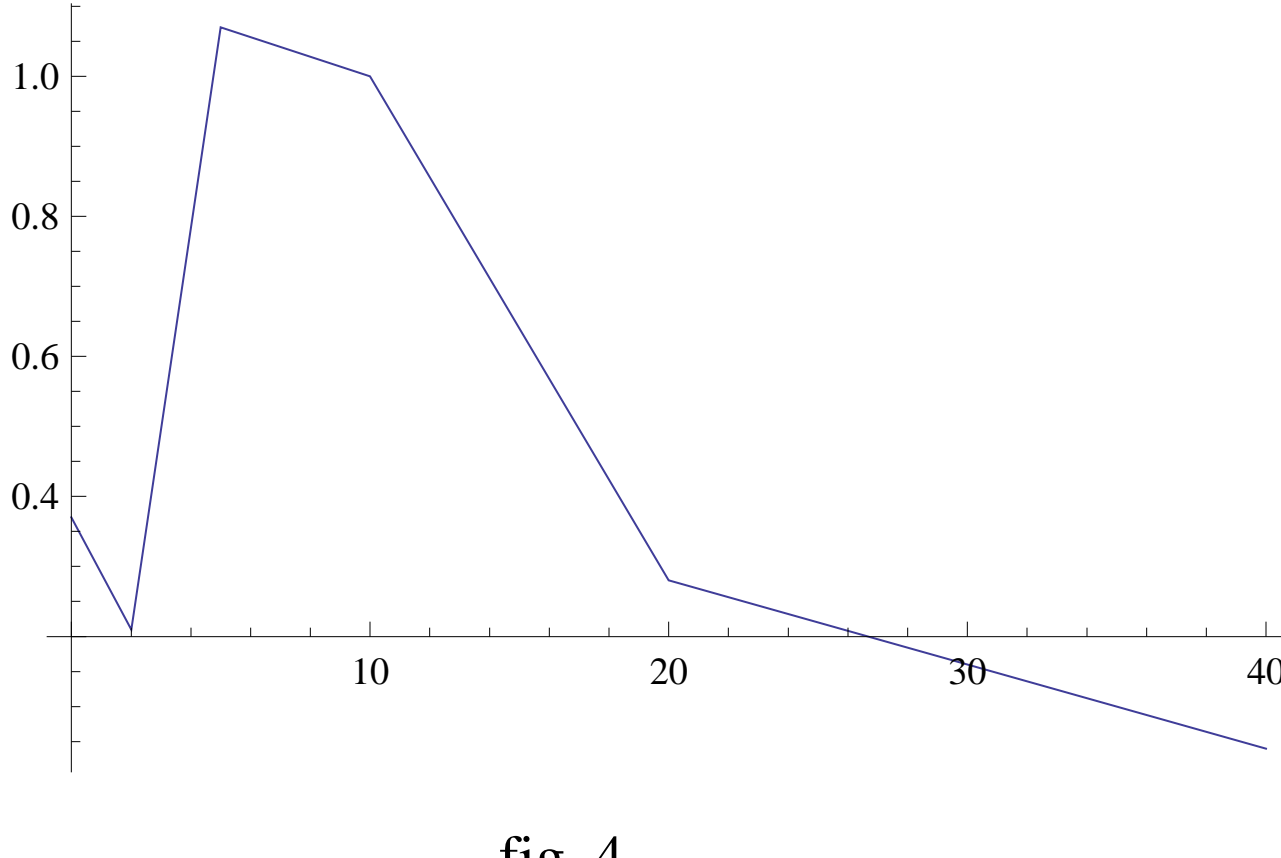

fig. 4

The calculated luminosity (W) of an $R_0$ = . 1 m, $Q_0$ = .0015 C ball as a function of time (sec).

For $Q_0$ = .0015 and $\alpha$ = .01 the ball is far less effective at heating its outside environment. It never achieves any noticeable brightness. If $\alpha$ is increased to 1 the ball rapidly dissipates its charge. It glows brightly but only over about 3 sec.

Consider the $Q_0$ = .0015 C case. Examining the electric field immediately beyond this ball's surface (at t = 0) we find that it is about $1.4 \times 10^9$ V/m – an enormous value ~3 orders of magnitude greater than the breakdown field strength for normal air. This condition would exist out to several meters beyond the ball. In some ways this is a good result – we are, at least, guaranteed a vigorous corona discharge to keep the ball hot. Of course, we do not think that the actual field inside the corona discharge is likely to be quite so high. It is, probably, screened somewhat by ions within the discharge itself. Owing to the strong field produced by the ball, there will always tend to be a net excess of oppositely charged ions close to the ball and more like-charged ions farther away. We have no reliable means of calculating just how much the effective field would be reduced. But it would have to still be quite large. (If it became too small there would no longer be a sufficient electric field to promote the creation of ions in the corona and this region would then fully re-experience the ball's unscreened field and ionize again.) This is problematic in that, were the ball to get too close to any object (e.g. a grounded conductor) onto which it could release its excess charge, the result would likely be a mini-lightning strike and the quick and violent destruction of our ball. This may explain the explosive deaths of some ball lightnings. No wonder ball lightning is so rare and fleeting – very favorable circumstances would have to exist for it to live out its 20 sec lifetime.

In light of the diminution of the effective field within the region of corona discharge our simulations might be looked upon as giving unrealistically long-lived solutions – being weaker, the real field would not allow the current flow to give up so much heat-energy. Not knowing exactly how to model the corona we can only look upon our numerical simulation as a kind of best-case scenario. We can also consider a worst-case scenario in which the effective E field outside the ball is so well-screened that it is at only about the breakdown field for air (which we will take to be $3 \times 10^6$ V/m) out to some rather large distance from the ball's surface. This field is only ca. $10^{-3}$ that assumed above. The energy produced by the escaping charges is, therefore, reduced by this factor. Hence $\alpha\,Q_0$ might have to be roughly 1000 times that employed before or the ball would cool much too quickly. As discussed above, there is only so large $\alpha$ can become. Assuming $\alpha = .1$ we find $Q_0 \approx 1$ C in this case; this is an unrealistic number, even given the total charge a lightning strike brings down. But we can calculate the results for a ball of this kind. We get a solution that is physically reasonable for as little as $Q_0$ = .03 C (fig. 5). This is because heat generation now drops off only as $\frac{1}{r^2}$. The ball expands to about twice its original size and lasts over 20 sec. Obviously, both scenarios are unrealistically extreme, and the latter one not strictly possible mathematically. The truth must lie somewhere in between.

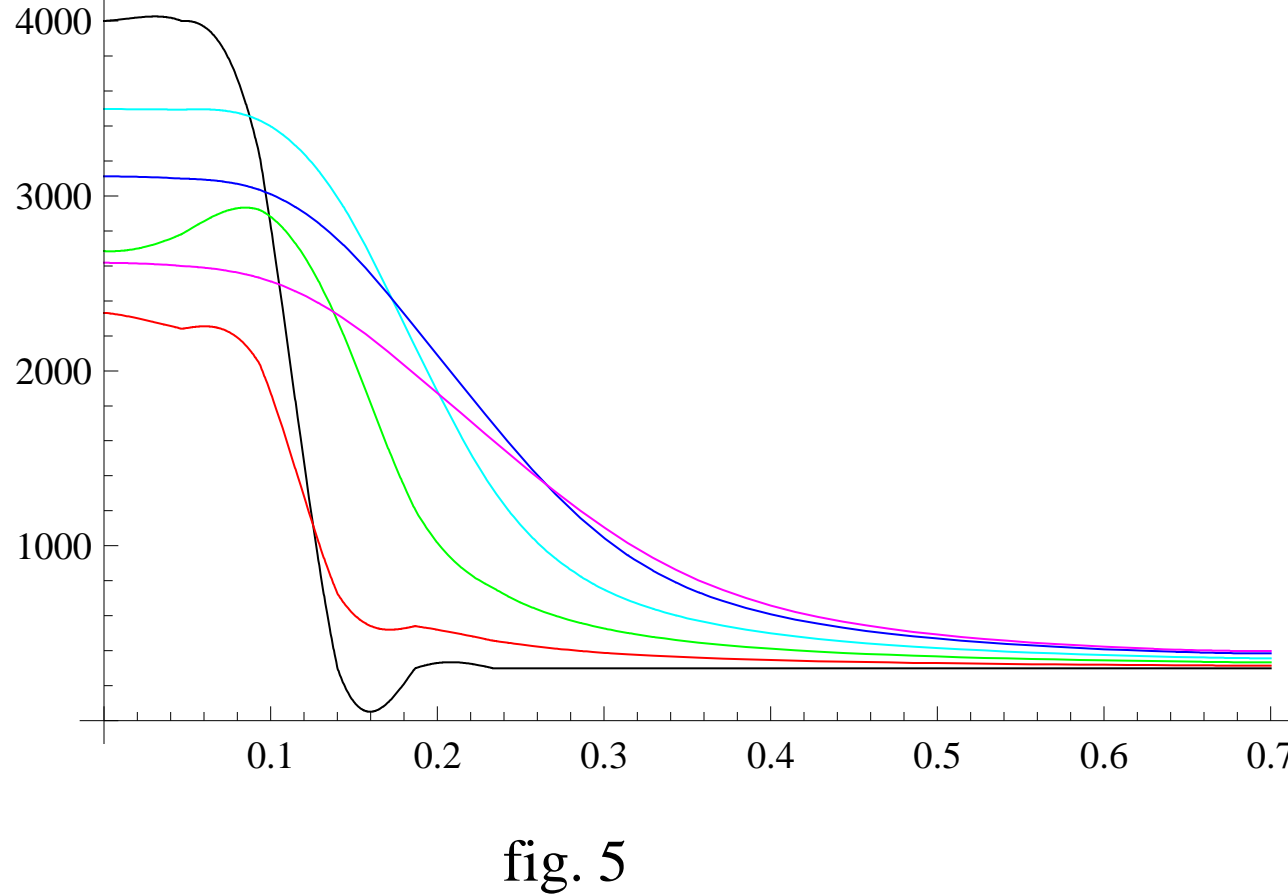

fig. 5

The calculated temperature profile ($K$) of an $R_0$ = . 1 m, $Q_0$ = .03 C, $\alpha$ = .1, ball with a constant external field as a function of radius (m) plotted for $t$ = 0 (black), 2 (red), 5 (green), 10 (blue-green), 20 (blue), 30 (purple) sec. This result was calculated using the original method.

This ball, having little mass, would feel an upward force of about 5X$10^{-2}$ N. But there could very well be ambient electric fields in its environment that might counteract this. The ball is a (highly) charged object and will be subject to these fields. Such balls as are observed are, necessarily, somewhat close to the ground. If a .0015 C ball attracted about 1.3X$10^{-8}$ C of charge in the ground below it would remain stable at about a height of 2 m. There might also be other E fields around that might help stabilize it. In any case, the ball's charge must be distributed uniformly throughout it; otherwise it would be torn apart. It may be objected that, as the ball's charge decays away, there will be less downward force holding it in place. True enough. But, while this process is occurring, the ball will generally be cooling and becoming less buoyant. Also, it will have had additional time to attract more charges into the ground beneath it. This may further help it not to convect away and dissipate. (Interestingly, the electric field at the Earth's surface, under normal circumstances, is ca. 100 V/m which is just about enough to keep the .0015 C ball in place assuming that it is positively charged.)

We must say that ball lightning occurs only under rather fortunate circumstances where the electric field in the surrounding environment balances off its natural buoyancy. But it is significant that we are, at least, able to say this. Not every model of ball lightning that proposes it to be hot can explain its not floating away. And the Cen spectra indicate clearly that these balls are hot indeed. We have to consider the possibility that the formation of hot, charged, masses of air is a rather common phenomenon occurring when lightning strikes down. Ball lightning would be uncommon only because unusual conditions are required for its survival. This possibility is not, necessarily, untestable. Perhaps what we see as ball lightning represents only the most energet-ically extreme manifestation of a more general phenomenon. Cooler, less highly charged, balls might well form even more easily and frequently. But these would be invisible to the naked eye. It is possible to trigger lightning strikes using model rockets attached to wires and attempts to create ball lightning in this way have been made (Hill et al. 2010). If the aftermath of such strikes could be filmed using highly sensitive night-vision or thermal-imaging equipment we might detect the frequent formation of such "sub-critical" balls.

Larger and smaller balls can be modeled in the same way. Equation 3) is non-linear and no simple scaling arguments can be made regarding its behavior – each case must be considered individually. We examine a ball with $R_0 = 1$ m. This is close to the largest size reported. If $Q_0 = 0$ the ball cools over about 90 sec. Interesting results are observed for $Q_0 = .1$ C (fig. 6). For this value it initially swells. Its luminosity is very great.

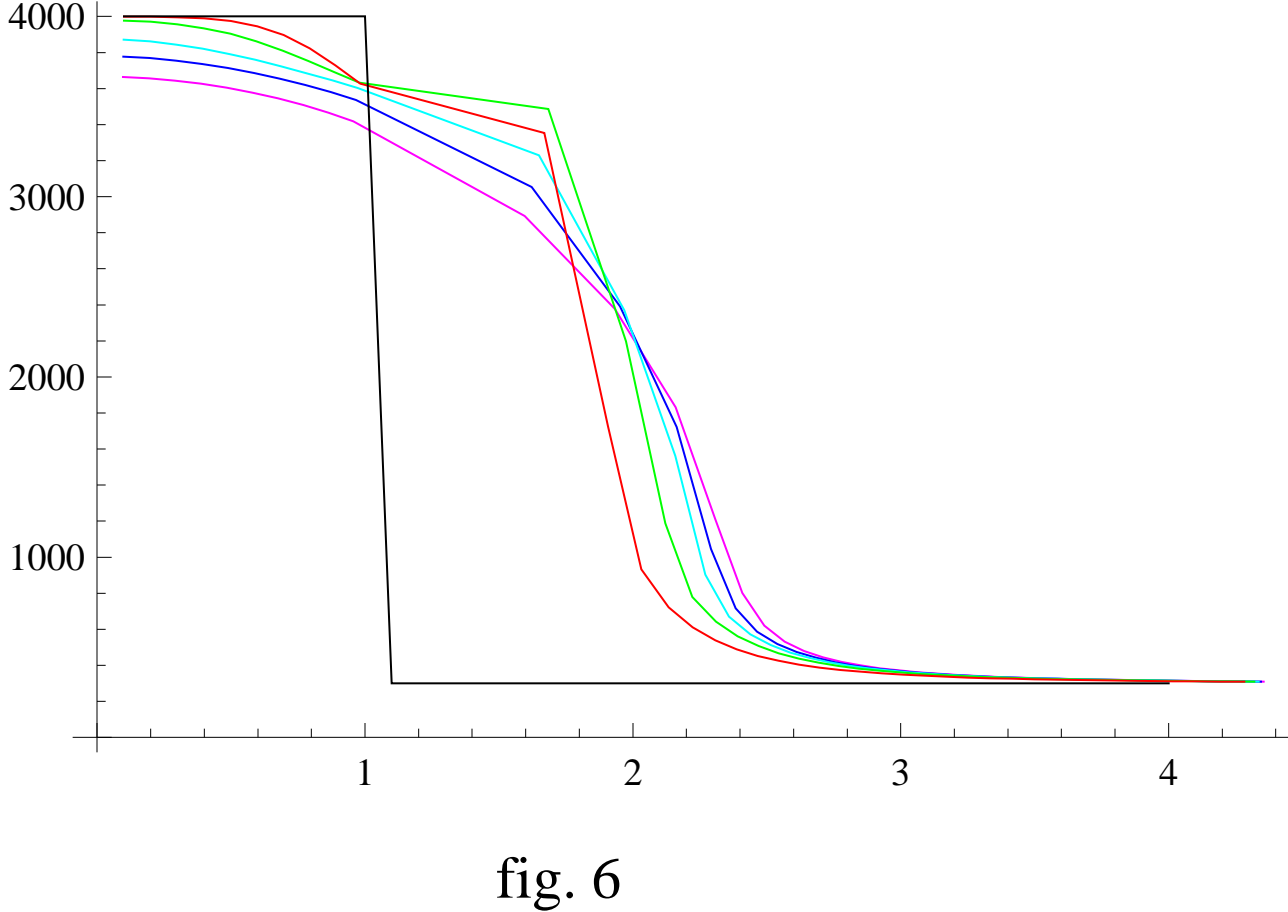

fig. 6

The calculated temperature profile ($K$) of an $R_0 = 1$ m, $Q_0 = .1$ C, $\alpha = .1$, ball as a function of radius (m) plotted for $t = 0$ (black), 10 (red), 20 (green), 40 (blue-green), 60 (blue), 90 (purple) sec.

An uncharged small ball with an initial radius of .01 m stays above $2000^o$ only about .02 sec. For $Q_0 =$ .00006 C it lives about 5 sec and achieves a maximum luminosity of ~ .003 W (fig. 7) at 2 sec. (See Supplementary Material *Mathematica*.) Its radius also swells quickly.

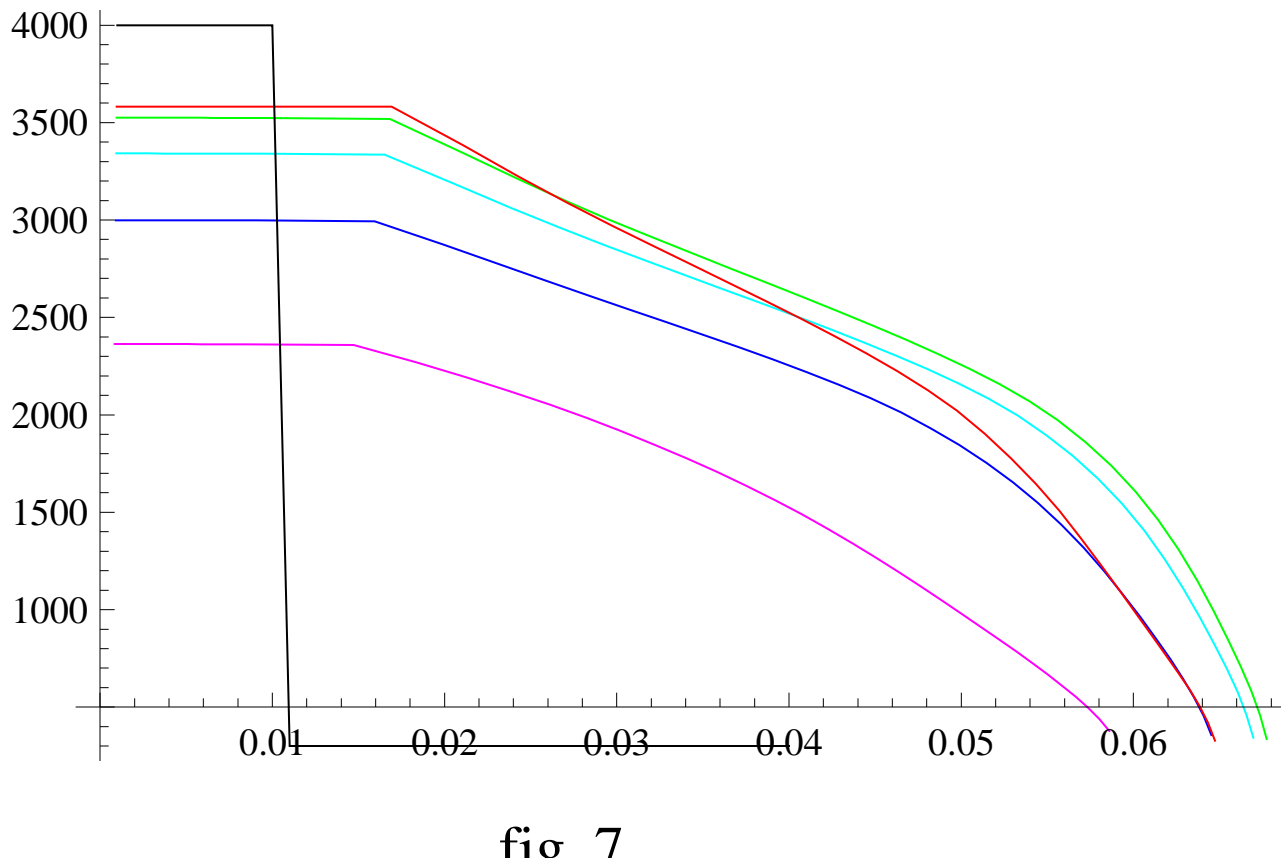

fig. 7

The calculated temperature profile ($K$) of an $R_0 =$ .01 m, $\alpha = .1$, $Q_0 = .00006$ C ball as a function of radius (m) plotted for $t = 0$ (black), 1 (red), 2 (green), 3 (blue-green), 5 (blue), 10 (purple) sec.

In passing from the original method of solution to our new method two important issues have been addressed. In the original paper *D(T (r, t))* was treated, locally, as a constant. But, more properly, the temperature flux should be expressed as $D(T(r, t)) \nabla T(r, t)$. The temperature will change as the divergence of this quantity. As long as the radial variation of the temperature is small, this correction makes little difference. But it is not always small everywhere. Also, originally, we tried to compensate for the expansion/contraction of the air as its temperature changes by using the method of (Christmann 1967). Here we do it in a much more straightforward way. It is, actually, rather surprising that our results are not changed very dramatically at all. The air outside the ball heats to a slightly greater extent. But, otherwise, things stay more-or-less the same.

## 4. Discussion.

If the charge of the ball is .0015 C there will be about $10^{16}$ charge carriers to begin with. It is, of course, quite arbitrary which particles we choose to call "charge carriers." They are just any ions or electrons that happen to be in excess. If the initial radius of the ball is .1 m, and if these are uniformly distributed and confined within this radius, the average distance between them and their nearest neighbors will be about $10^{-6}$ m. If the temperature in this region is about $4,000\,K$ the oxygen molecules (which we concentrate on since $O_2$ has the lowest ionization energy) are about $10^{-4}$ % ionized (as follows from the Saha equation). The predominant charged species will be ${O_2}^+$ and electrons (Bauer 1990). The Debye length is therefore about $5X10^{-6}$ m. It is reasonable to invoke this mechanism to the extent that the number density of screening ions in the environment greatly exceeds that of the 'charge carriers.' Here they are about the same ($6X10^{17}$ and $2.5X10^{18}$ $m^{-3}$, respectively). Obviously, we cannot invoke our electrostatic screening mechanism here; all the charge carriers would be rapidly expelled from the ball. If, however, we imagine that the air inside the ball contains even a very small number of molecules with low ionization energies the number density of 'screening particles' will be increased greatly. The effect of such contaminants on uncharged balls has been studied by Lowke et. al. (1969). Any highly ionizable salt would suffice. But we will use NaCl as an example since it is a common substance and its chemistry is well-understood. Suppose it is present to the extent of 0.1%. NaCl begins to ionize as it enters its gaseous state. Its boiling point is 1738 K. Above 3000 K we would expect it to be very highly ionized. The added NaCl will contribute a number density of $1.8X10^{21}$ $m^{-3}$of both sodium and chloride ions to the pool of screening particles (which is roughly 1000X that of the charge carriers). Well-screened, the charge carriers will only feel repulsive forces from other charges very nearby them. And these forces might just as easily push them towards the center of the ball as outwards. As long as the charge carriers, however defined, remain well-screened the charge will not all immediately fly away to infinity. It enjoys a natural mechanism of containment.

Positively charged ions have small mean free paths ($\sim 7X10^{-8}$ m) and are hugely massive – they just cannot go very far by diffusion. A nitrogen molecule in air at STP travels about 7 mm in 1 sec and 21 mm in 10 sec. They will go a bit farther in hotter air but this will not alter our conclusions very drastically; by the time these charge carriers move very far in relation to the ball's size most of the excess charge will have been expended anyway and heat production will have largely ceased. We do not much care what happens to them afterwards. It is harder to estimate the mean free path for electrons but it is generally at least 100 times that of ions and molecules. If the excess charge carriers are electrons they will quickly diffuse into the region where the electric field is largely unscreened and be propelled away to infinity. This would necessarily result in a situation where $\alpha$ would be quite large. This is not compatible with our mechanism, as has been noted above. If the ball is positively charged the electrons will be attracted back towards it. We, therefore, conclude that the ball is *positively charged.*

The high field strength around the ball is difficult to understand. We have already observed that it may be screened somewhat. But it would have to be quite large for the ball to sustain itself according to our mechanism. Very large electric fields, some probably exceeding $3X10^6$ V/m, undoubtedly exist between thunderclouds and the ground. But the entire sky does not break down instantaneously. The circumstances around our ball may be similar – its quasi-stable state only exists as long as it cannot find anything onto, or any way in which, which to release its charge. When it does, it dies explosively. (A .0015 C ball, were it to explode in this way, would produce an energy more-or-less comparable to that of two M80 firecrackers.) In the meantime it releases what charge it can through a positive corona discharge. And for a positively charged ball to explode it has to be hit by a mini-lightning strike coming from a source of electrons that may be far away where the field is small. This process may have a more difficult time occurring than were the ball negatively charged; then the electrons would already be in an area where the field was intense.

It is, also, rather easier to understand the formation of a positively charged ball. If the initial mass of hot, ionized, air were to find itself around some positively charged object some of its electrons, which have great mobility, might be drawn away, thus leaving it with a net positive charge. We could even imagine something like a very brief mini-lightning strike carrying away the electrons by providing a conducting path between the nascent ball and the object.

## 5. Conclusion.

The strongest argument in favor of the above-described model is its ability to produce results that agree fairly well with what has been reported in spite of being rather inflexible – it has only three adjustable parameters, $R_0$, $Q_0$ and $\alpha$. And its agreeable results only come about when these parameters are in what would seem to be a physically reasonable range. Everything else is determined by the relatively straightforward physics of the model and by numbers that can be looked up in, or extrapolated from, the literature. Also, it does not require any external source of power; its energy comes only from the heat and charge that were in the lightning channel that created the ball. Moreover, it provides an explanation for the spherical shape of the phenomenon. The ball's charge provides a mechanism whereby its natural buoyancy may be counteracted. The presence of NI and OI lines in the Cen ball lightning spectra is consistent with this model (although it is consistent with some other models as well). This model also explains the lack of a sharply-defined boundary to the ball – something eyewitnesses have mentioned. It explains the reported hissing and corona discharges associated with the phenomenon – something very few other models do. It explains why some people have been shocked, and even killed, by the phenomenon and why it is frequently observed to explode. It is also important to point out that we are able to provide a rough mechanism whereby the ball's excess charge is, mostly, contained within its initial radius.

A shortcoming of this model is its inability to predict $\alpha$ – there is, admittedly, something circular about assuming that $\alpha = .1$ $\text{sec}^{-1}$ because ball lightning lasts about 10 sec. But, still, it is significant that we get physically plausible solutions for $\alpha$ in this range; if the model only worked for $\alpha = 10^7$ or $10^{-15}$ $\text{sec}^{-1}$ we would have to say it was nonsense. We note that, if the charge dissipated solely through the passive diffusion of positive ions out of the ball, the ball lives longer than predicted above (but is still of the same order of magnitude). We conclude that most of the charge dissipation comes about by ionization events in the corona discharge. The electrons so produced are drawn into the ball and the positive ions are expelled to infinity.

The surprisingly high electric field strength immediately outside the ball is somewhat perplexing. Knowing of no reliable way to simulate such a situation, we have contented ourselves with a simple model that deals only with the net energetics of each unit of charge that leaves the ball and departs to infinity. We pay no attention to the details of its departure. And the assumption that the ball's charge decays exponentially with time is, itself, arbitrary. But this is not really a bad assumption. Regardless of the details of the corona discharge process, we know two things with certainty: The ball's charge will decay away and it will do so over some fairly definite period of time (which we may as well call $1/\alpha$). Modeling this process crudely as $Q(t) = Q_0 e^{-\alpha t}$, or trying to simulate $Q(t)$ more realistically, might not produce very different or more reliable answers given the approximate way in which we approach the problem and in which computers solve PDEs.

## Acknowledgement.

The author is grateful to Professor M. A. Uman (U. of Florida) for the Lowke reference and many useful discussions and to Marek and Robert Nelson (PDE Solutions Inc.) for help with FlexPDE.

## References.

**(Since ArXiv does not support the posting of multiple pdf. files I have simply attached the Supplementary Material. This file is also available as a Mathematica notebook. It could easily be run by any user having *Mathematica* on his computer. It is readily obtainable by contacting the author at the above email address.)**

## Supplementary Material Mathematica.

**Below we define W[T], $D_0$ {T}, $\gamma$[T] and $\beta$[T]. Our values are taken from (Lowke, J.J., M.A.Uman, R.W.Liebermann, 1969 : Toward a theory of ball lightning.*J. Geophys. Res.*, 74, 6887 - 6898.) , (Yos, J.M., 1963 : Transport properties of nitrogen, hydrogen, oxygen, and air up to 30 000 K.*AVCO Tech. Memo*.RAD - TM - 65 - 7, AD - 486 068, doi : http : // www.dtic.mil / dtic / tr / fulltext / u2 / 435 053. pdf.), and engineeringToolbox.com.**

```
W[t_] := Exp[.00115 t + 1.5567]
```

```
D0 = Interpolation[{{300, .000019}, {1000, .00015}, {2000, .00052}, {3000, .0024},
   {4000, .0043}, {5000, .009}, {6000, .027}, {7000, .053}, {10 000, .024}}]
```

```
InterpolatingFunction[{{300., 10 000.}}, <>]
```

```
L = Interpolation[{{300, 0}, {1000, 6.6 × 10 ^ -7},
   {2000, .005}, {3000, .06}, {4000, .19}, {5000, .34}, {6000, .48}}]
```

```
InterpolatingFunction[{{300., 6000.}}, <>]
```

```
γ = Interpolation[
  {{200, 1.406}, {300, 1.402}, {500, 1.387}, {1100, 1.329}, {1900, 1.301}, {6000, 1.286}}]
```

```
InterpolatingFunction[{{200., 6000.}}, <>]
```

```
β = Interpolation[
  {{200, 2.46}, {300, 2.49}, {500, 2.58}, {1100, 3.04}, {1900, 3.32}, {6000, 3.5}}]
```

```
InterpolatingFunction[{{200., 6000.}}, <>]
```

**Below we define our physical constants :**

```
k = 1.38 × 10 ^ -23
P0 = 101 325
h = 10 ^ -34
a = .1
e0 = 8.85 × 10 ^ -12
Q = .0015
```

**Below we define the Laplacian of T :**

```
HH[U_] :=
 Table[{i, Which[i == 1, 2 ( U[[2]][[2]] - U[[1]][[2]]) / U[[1]][[1]] ^ 2, i == 40, 0, i == 41, 0,
    True, 2 ((U[[i + 1]][[2]] - U[[i]][[2]]) / (U[[i + 1]][[1]] - U[[i]][[1]]) -
         (U[[i]][[2]] - U[[i - 1]][[2]]) / (U[[i]][[1]] - U[[i - 1]][[1]])) /
       (U[[i + 1]][[1]] - U[[i - 1]][[1]]) + 2 (U[[i]][[2]] - U[[i - 1]][[2]]) /
       ((U[[i]][[1]] - U[[i - 1]][[1]]) U[[i]][[1]])]}, {i, 1, 40}]
```

```
G[U_, M0_, i_] := (U[[i]][[2]] / M0[[i]][[2]])
```

**Below is Grad[$D_0$ ] · Grad[T] :**

```
H[U_] := Table[{i, Which[i == 1, 0, i == 40, 0, i == 41, 0, True,
    (D0[U[[i]][[2]]] - D0[U[[i - 1]][[2]]]) / (U[[i]][[1]] - U[[i - 1]][[1]])
     (U[[i]][[2]] - U[[i - 1]][[2]]) / (U[[i]][[1]] - U[[i - 1]][[1]])]}, {i, 1, 40}]
```

**This diffuses the temperatures :**

```
Diff[U_] := Table[If[i ≤ 40,
   {U[[i]][[1]], U[[i]][[2]] + .001 D0[U[[i]][[2]]] HH[U][[i]][[2]] + .001 H[U][[i]][[2]]},
   U[[i]] + .001 {0, 1}], {i, 1, 41}]
```

**Here is the initial state :**

```
K = Table[Which[i ≤ 10, {.01 i, 4000}, 10 < i ≤ 40, {.01 i, 300}, True, {0, 0}], {i, 1, 41}]
```

```
{{0.01, 4000}, {0.02, 4000}, {0.03, 4000}, {0.04, 4000}, {0.05, 4000}, {0.06, 4000},
 {0.07, 4000}, {0.08, 4000}, {0.09, 4000}, {0.1, 4000}, {0.11, 300}, {0.12, 300}, {0.13, 300},
 {0.14, 300}, {0.15, 300}, {0.16, 300}, {0.17, 300}, {0.18, 300}, {0.19, 300}, {0.2, 300},
 {0.21, 300}, {0.22, 300}, {0.23, 300}, {0.24, 300}, {0.25, 300}, {0.26, 300}, {0.27, 300},
 {0.28, 300}, {0.29, 300}, {0.3, 300}, {0.31, 300}, {0.32, 300}, {0.33, 300}, {0.34, 300},
 {0.35, 300}, {0.36, 300}, {0.37, 300}, {0.38, 300}, {0.39, 300}, {0.4, 300}, {0, 0}}
```

**Below we define the Source term :**

```
a Exp[-2 a t] / (16 Pi^2 r^4 β1 P0 e0)
```

$$\frac{706.188\, e^{-0.2\, t}}{r^4\, \beta 1}$$

```
Source[U_] :=
 Table[Which[i ≤ 10, U[[i]], 10 < i < 41, .001 {0, (706 Q^2 e^(-0.2` U[[41]][[2]])) / (β[U[[i]][[2]]] U[[i]][[1]]^4) U[[i]][[2]]} +
    U[[i]] - {0, 1} .001 U[[i]][[2]] W[U[[i]][[2]]] / (P0 β[U[[i]][[2]]]),
   True, {0, 0} + U[[i]]], {i, 1, 41}]
```

```
HD[U_] := Diff[Source[U]]
```

**FFF resizes the shells adiabatically :**

```
FFF[U_, M0_] := Table[If[j ≤ 40,
   {(Sum_{k=1}^{j} (G[U, M0, k]^(1/γ[U[[k]][[2]]]) (U[[k]][[1]]^3 - If[k > 1, 1, 0] U[[k - 1]][[1]]^3)))^
     (1/3), U[[j]][[2]] G[U, M0, j]^((1/γ[U[[j]][[2]]]) - 1)}, U[[41]]], {j, 1, 41}]
```

```
Evolve[U_] := FFF[HD[U], U]
```

```
NestList[Evolve, K, 45 001];
```

**Here we plot the core temperature :**

```
CT[U_] := U[[1]][[2]]
```

```
Table[CT[%35[[i]]], {i, 1, 45 001}];
```

```
ListPlot[%39, PlotRange → All, PlotStyle → RGBColor[1, 0, 0]]
```

```
ListPlot[Drop[%47, -1], Joined → True, PlotStyle → Black]
```

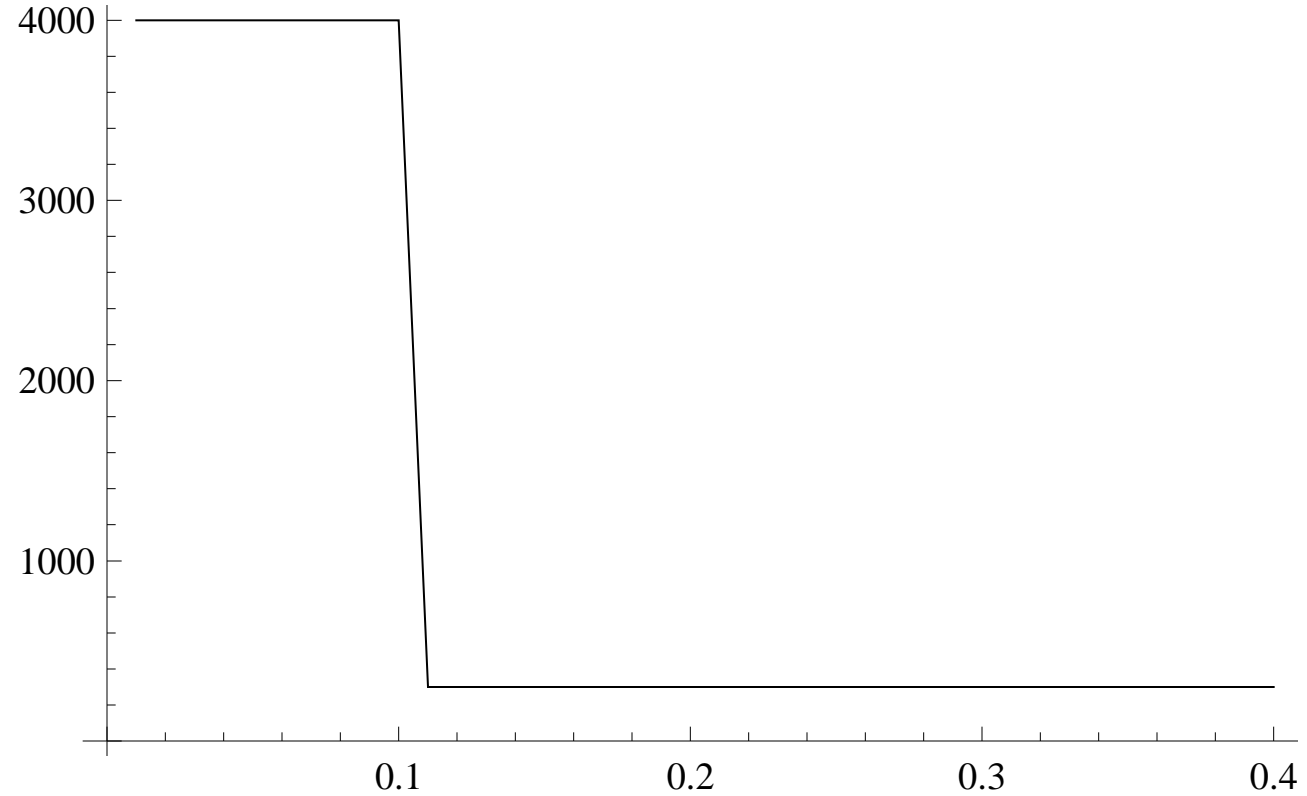

```
ListPlot[Drop[%42, -1], Joined → True, PlotStyle → RGBColor[1, 0, 0]]
```

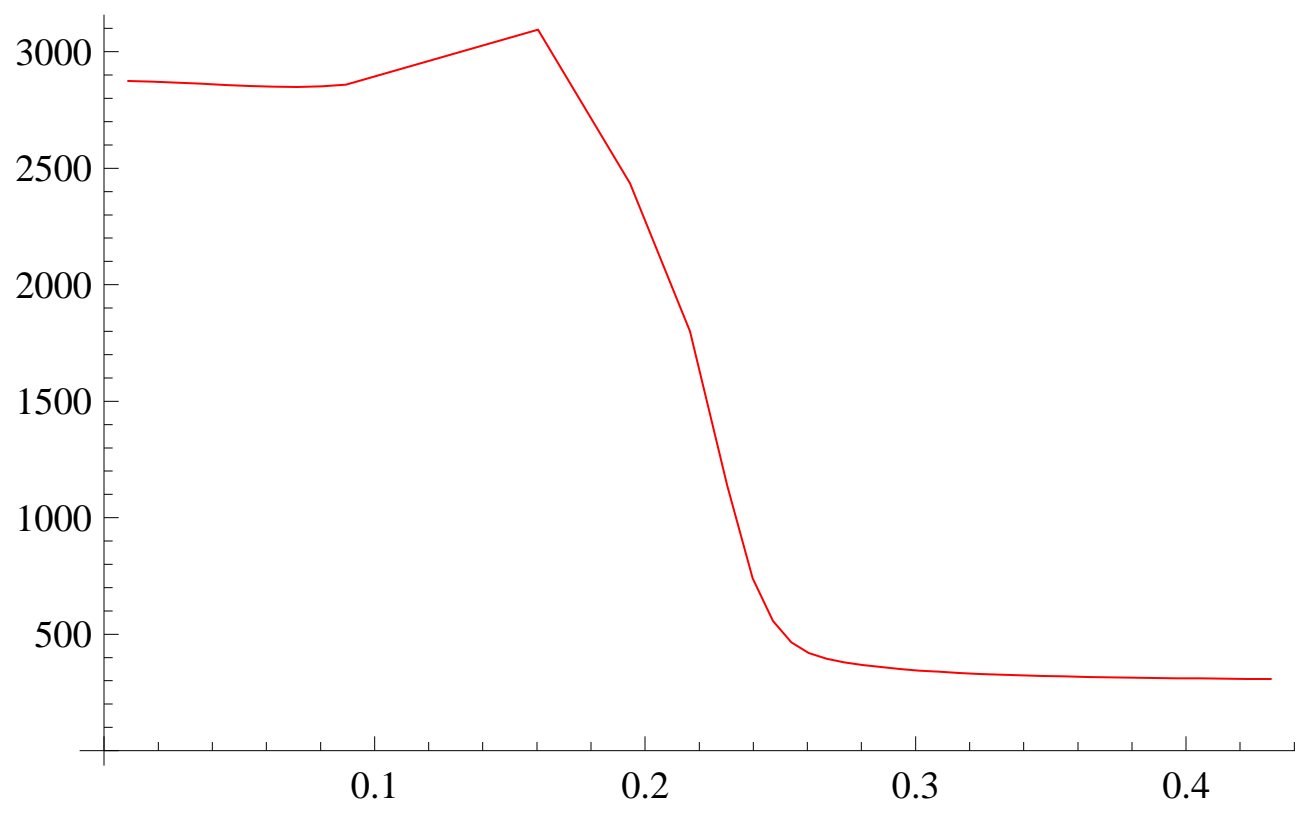

```
ListPlot[Drop[%43, -1], Joined → True, PlotStyle → RGBColor[0, 1, 0]]
```

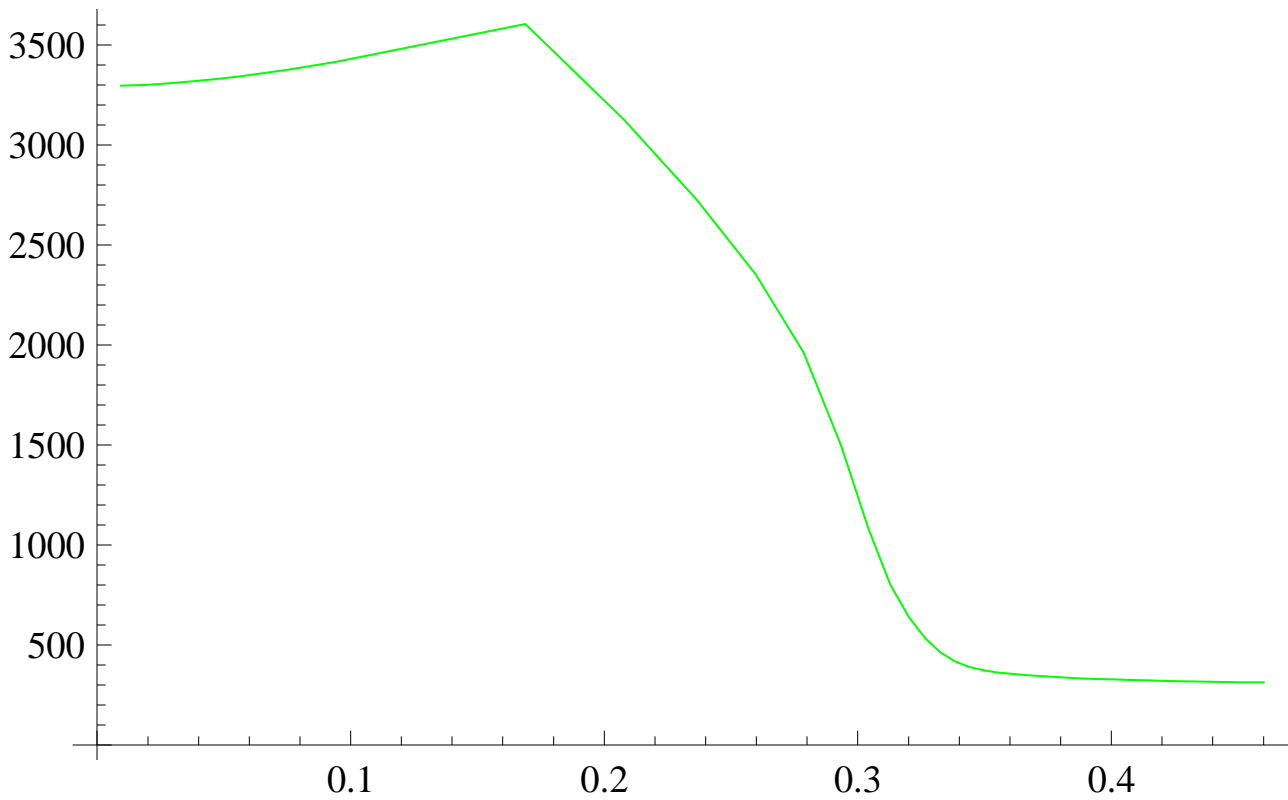

```
ListPlot[Drop[%44, -1], Joined → True, PlotStyle → RGBColor[0, 1, 1]]
```

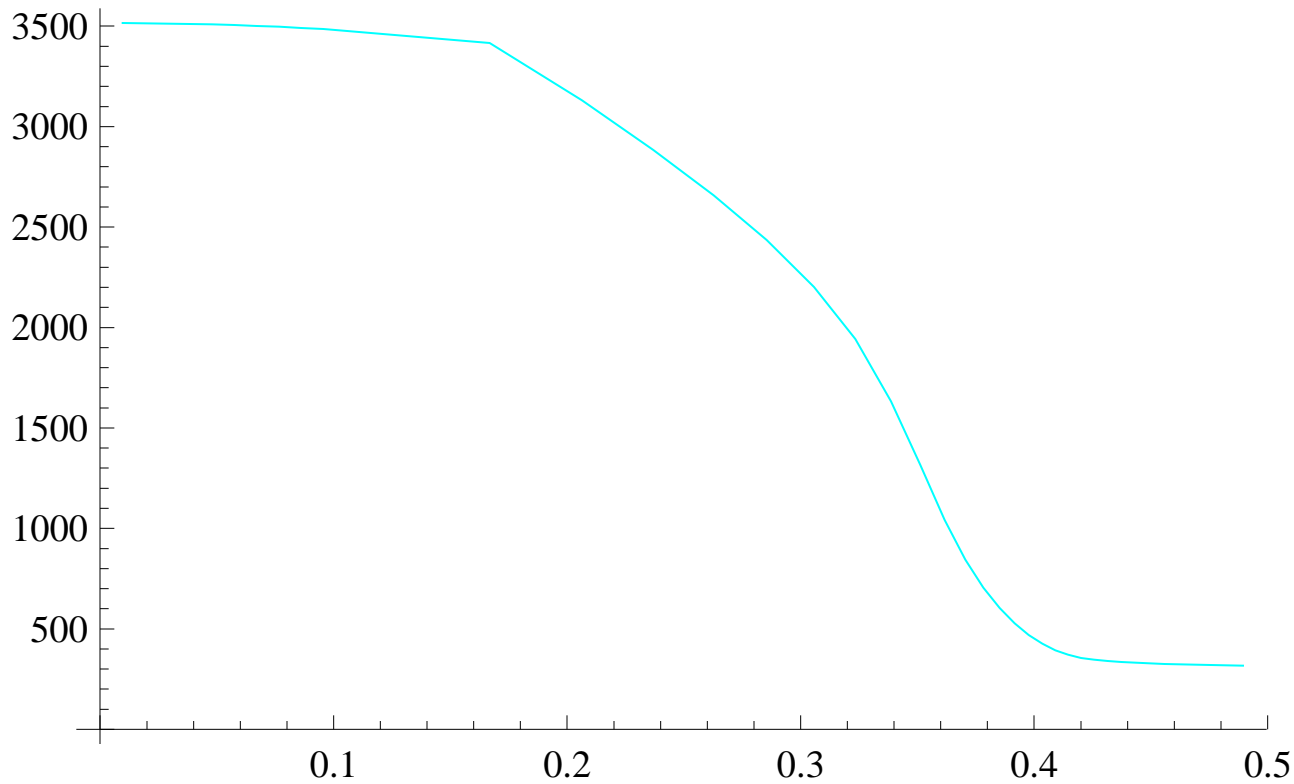

```
ListPlot[Drop[%45, -1], Joined → True, PlotStyle → RGBColor[0, 0, 1]]
```

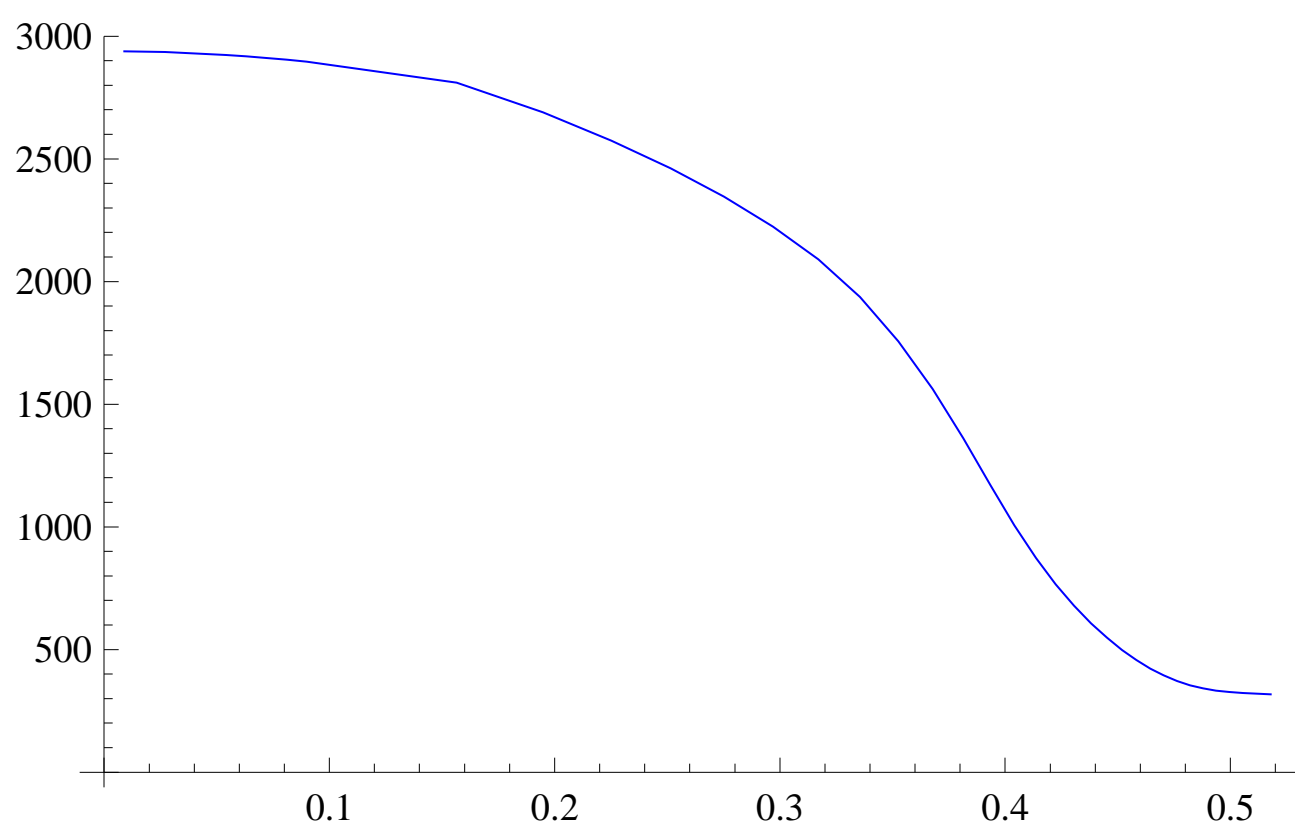

```
ListPlot[Drop[%48, -1], Joined → True, PlotStyle → RGBColor[1, 0, 1]]
```

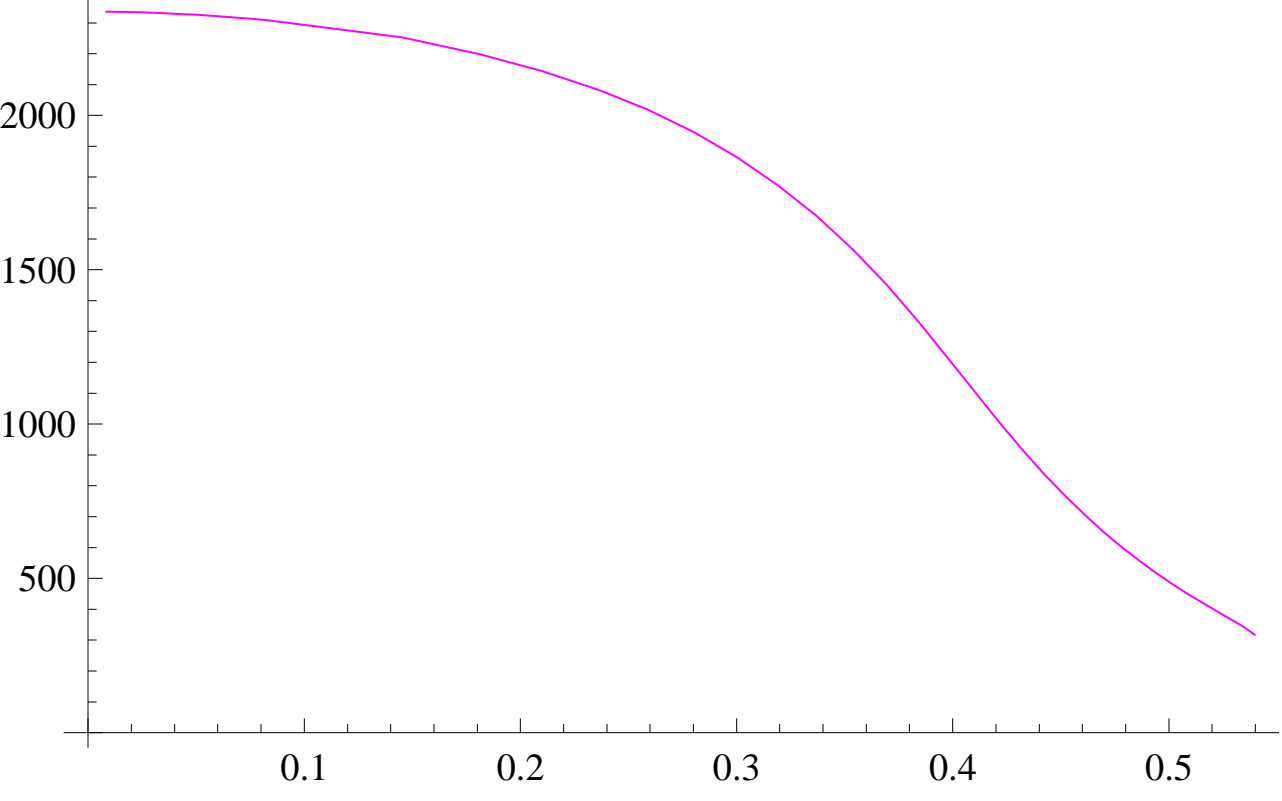

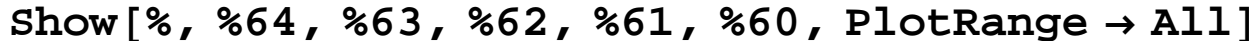

```
Show[%, %64, %63, %62, %61, %60, PlotRange → All]
```

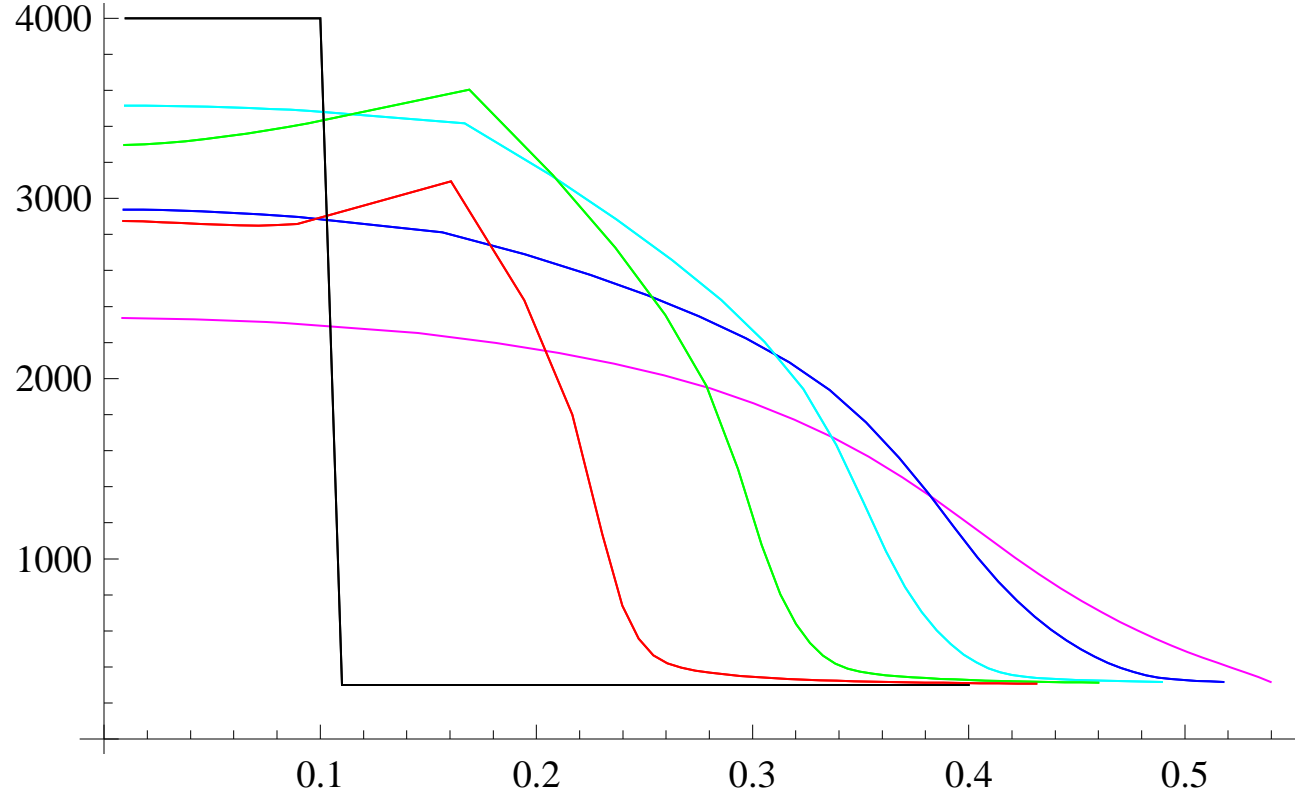

**Here we calculate the net luminosity of the ball. We sum the energy output of each shell after multiplying it by L which represents the fraction of energy released in the visible spectrum (assuming, very crudely, that the shells behave like blackbodies).**

```
TT[{x_, y_}] := y
```

```
J[U_] :=
 (4 Pi / 3) Table[If[i == 1, U[[i]][[1]]^3, (U[[i]][[1]]^3 - U[[i - 1]][[1]]^3)], {i, 1, 40}]
```

```
Lum[U_] := ((Map[W, Drop[Map[TT, U], -1]]) (J[U])).Map[L, Drop[Map[TT, U], -1]]
```

```
Lum[%90]
```

```
0.375546
```

```
Lum[%91]
```

```
0.208231
```

```
Lum[%92]
```

```
1.0705
```

```
Lum[%93]
```

```
0.998833
```

```
Lum[%94]
```

```
0.282508
```

```
Lum[%95]
```

```
0.0445447
```

```
{{0, .37}, {2, .21}, {5, 1.07}, {10, 1}, {20, .28}, {40, .04}}
```

```
{{0, 0.37}, {2, 0.21}, {5, 1.07}, {10, 1}, {20, 0.28}, {40, 0.04}}
```

**Here we look at an uncharged ball :**

```
Q = 0
```

```
NestList[Evolve, K, 45 001];
```

```
Table[CT[%29[[i]]], {i, 1, 45 001}];
```

```
ListPlot[%, PlotRange → All]
```

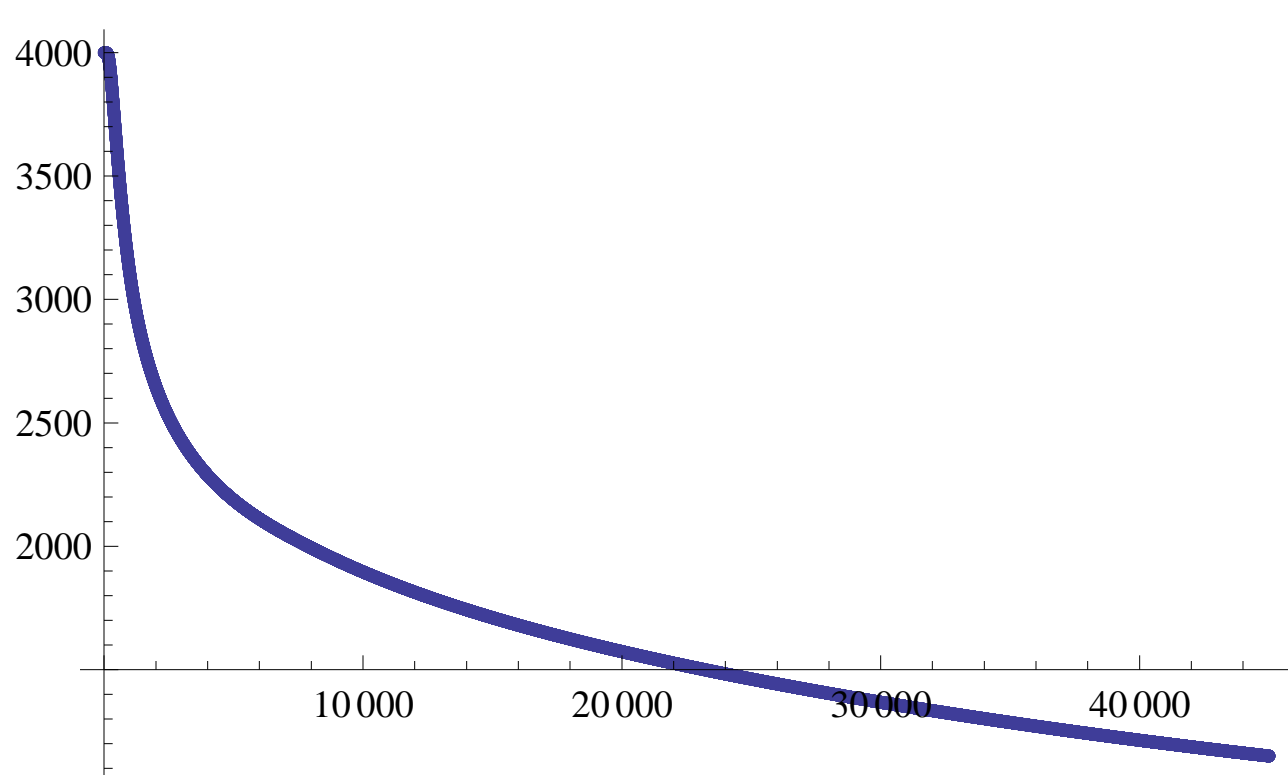

```
Show[%33, %32]
```

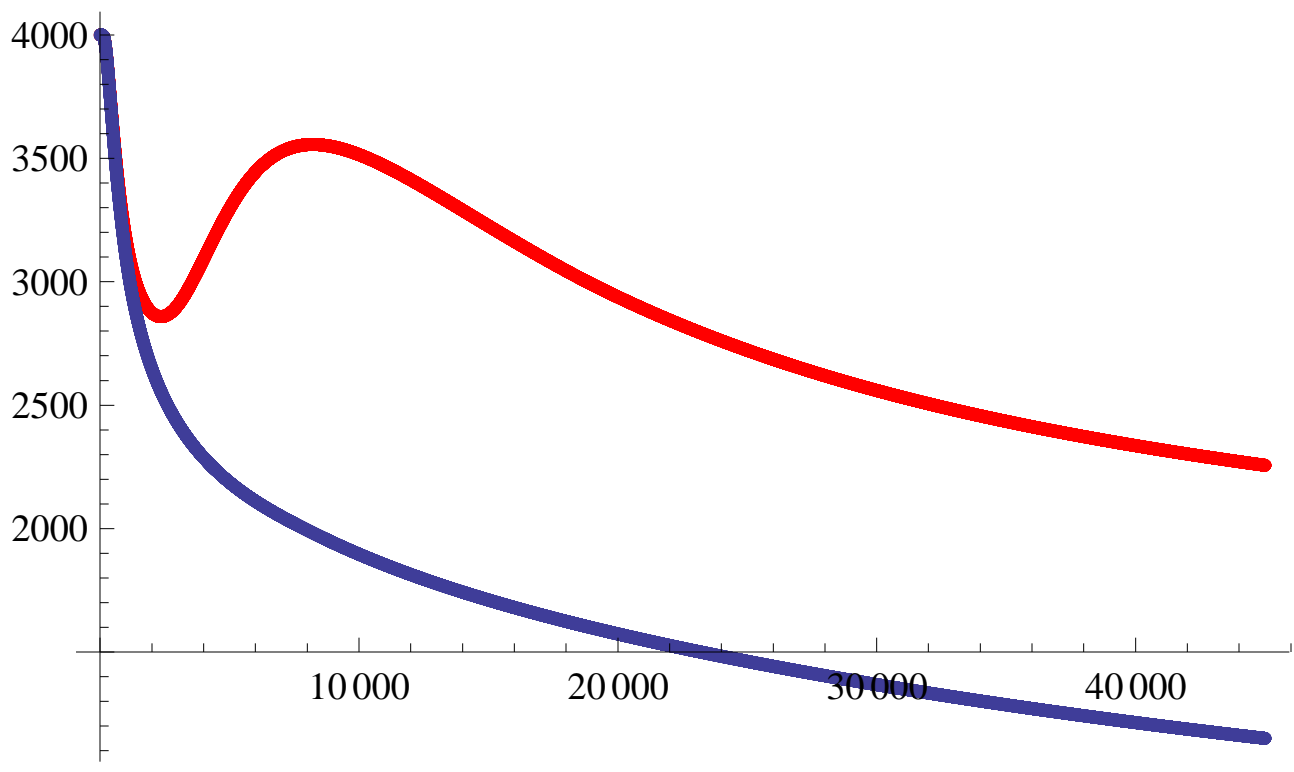

```
ListPlot[Drop[%35, -1], Joined → True, PlotStyle → Black]
```

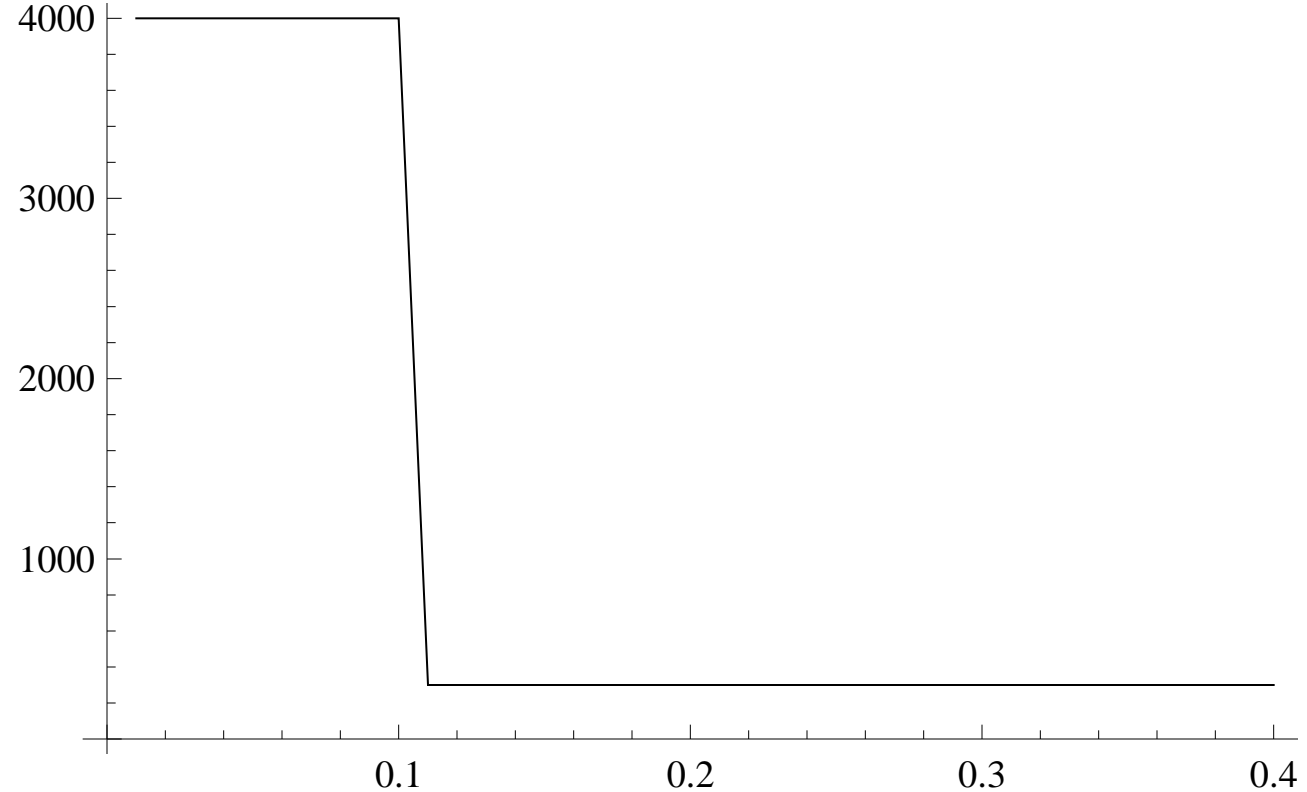

```
ListPlot[Drop[%36, -1], Joined → True, PlotStyle → RGBColor[1, 0, 0]]
```

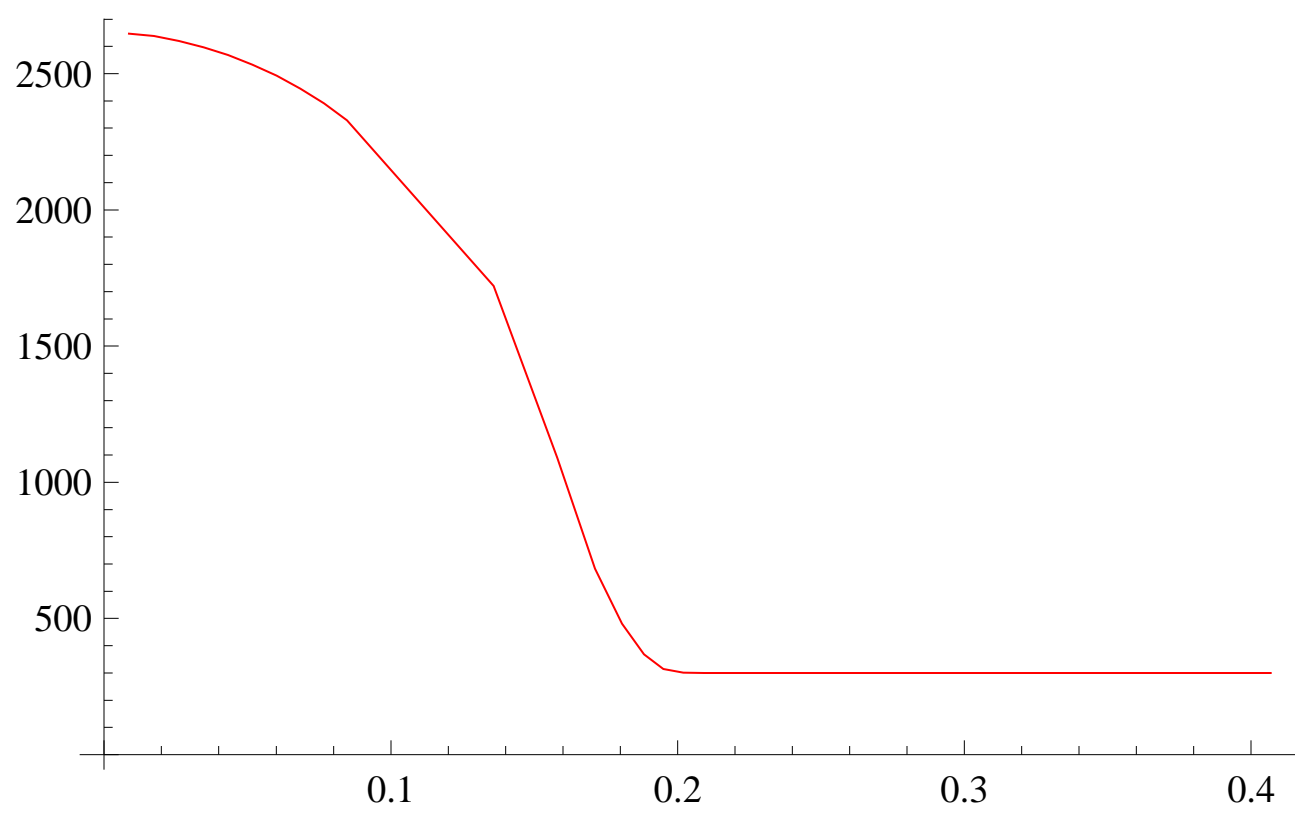

```
ListPlot[Drop[%37, -1], Joined → True, PlotStyle → RGBColor[0, 1, 0]]
```

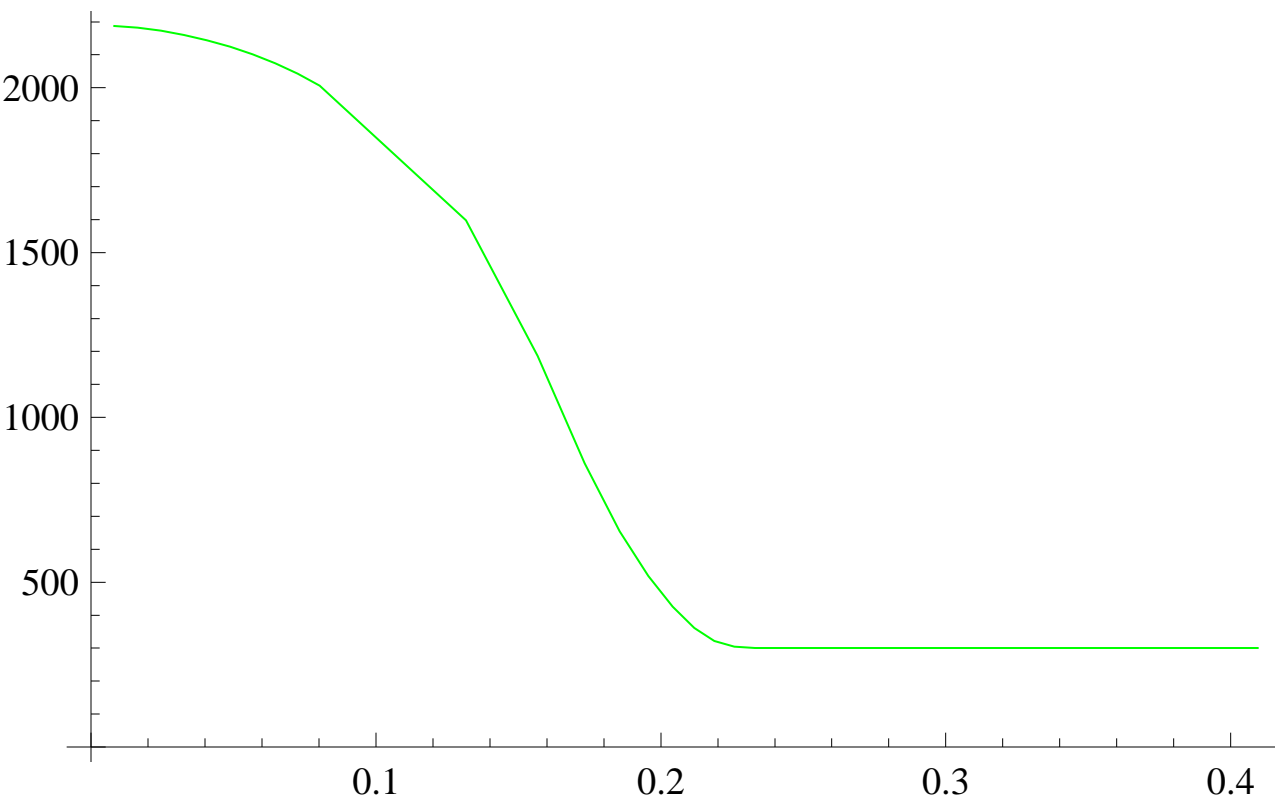

```
ListPlot[Drop[%38, -1], Joined → True, PlotStyle → RGBColor[0, 1, 1]]
```

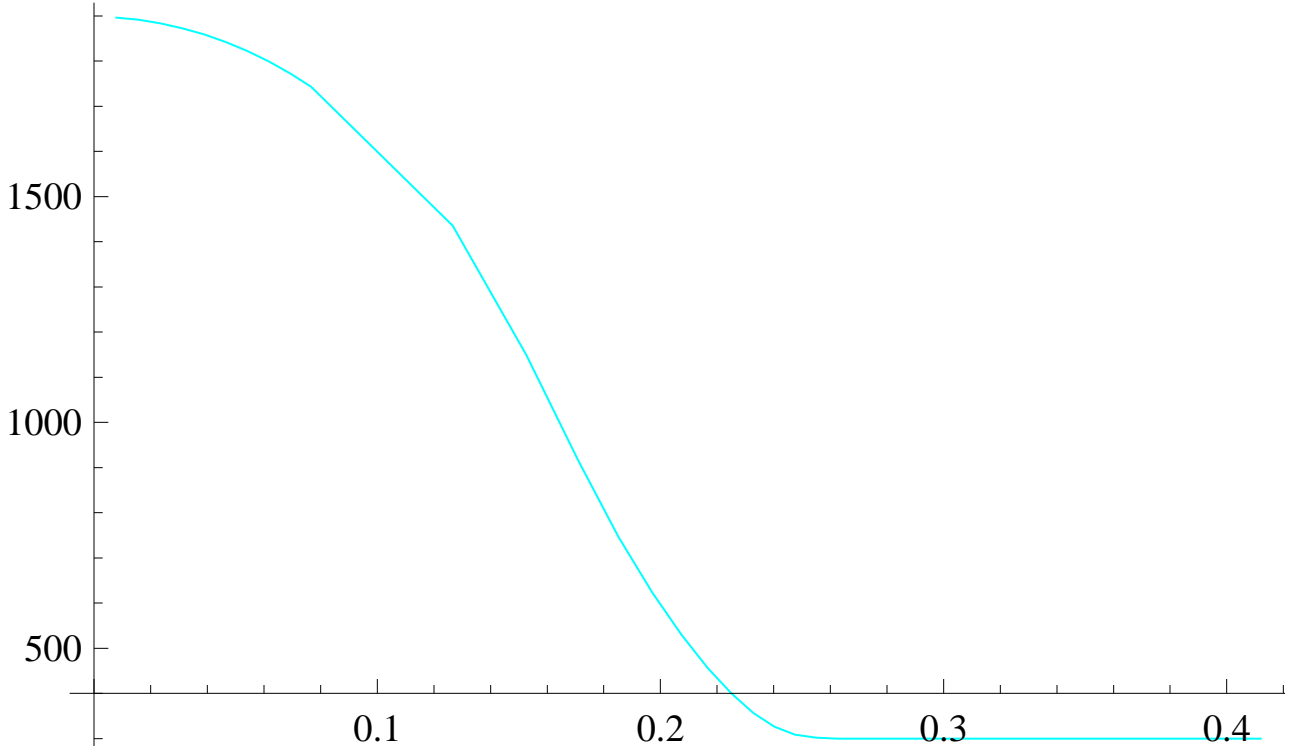

```
ListPlot[Drop[%40, -1], Joined → True, PlotStyle → RGBColor[0, 0, 1]]
```

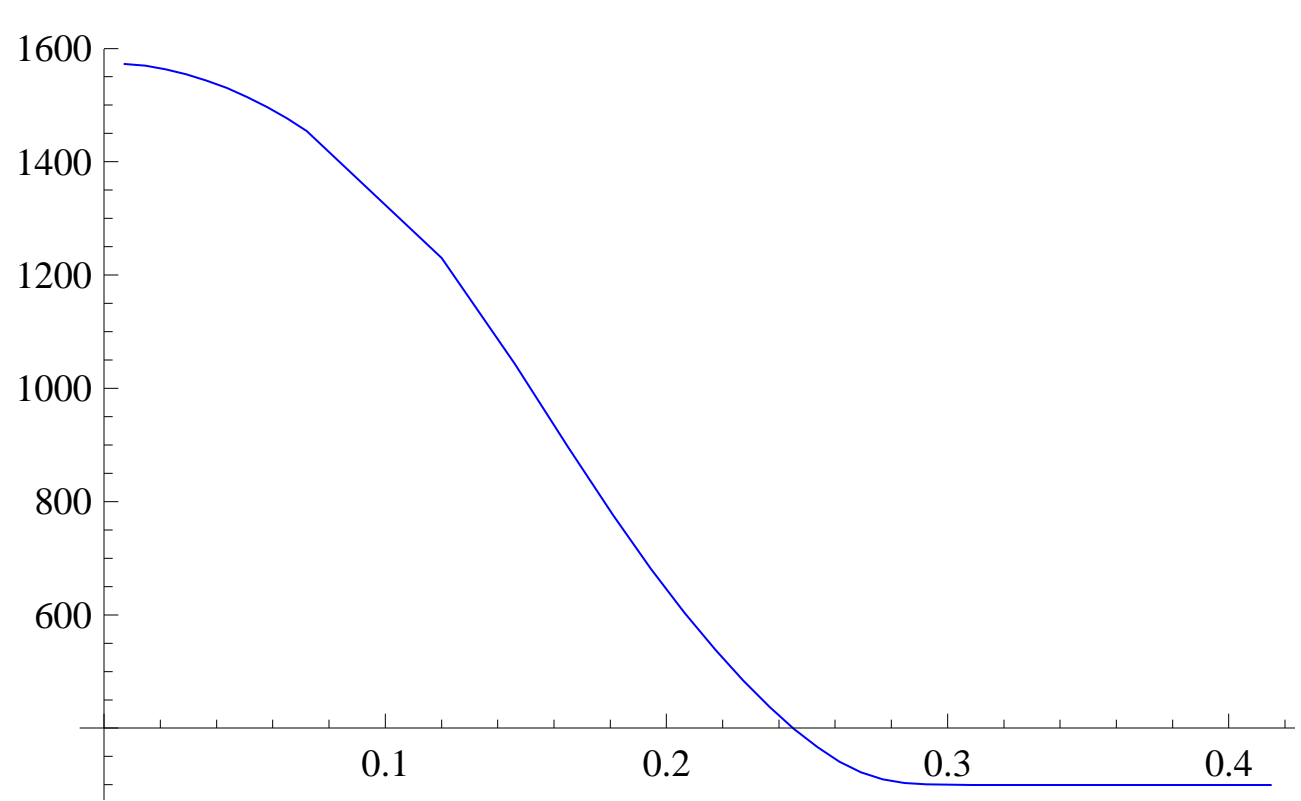

```
ListPlot[Drop[%41, -1], Joined → True, PlotStyle → RGBColor[1, 0, 1]]
```

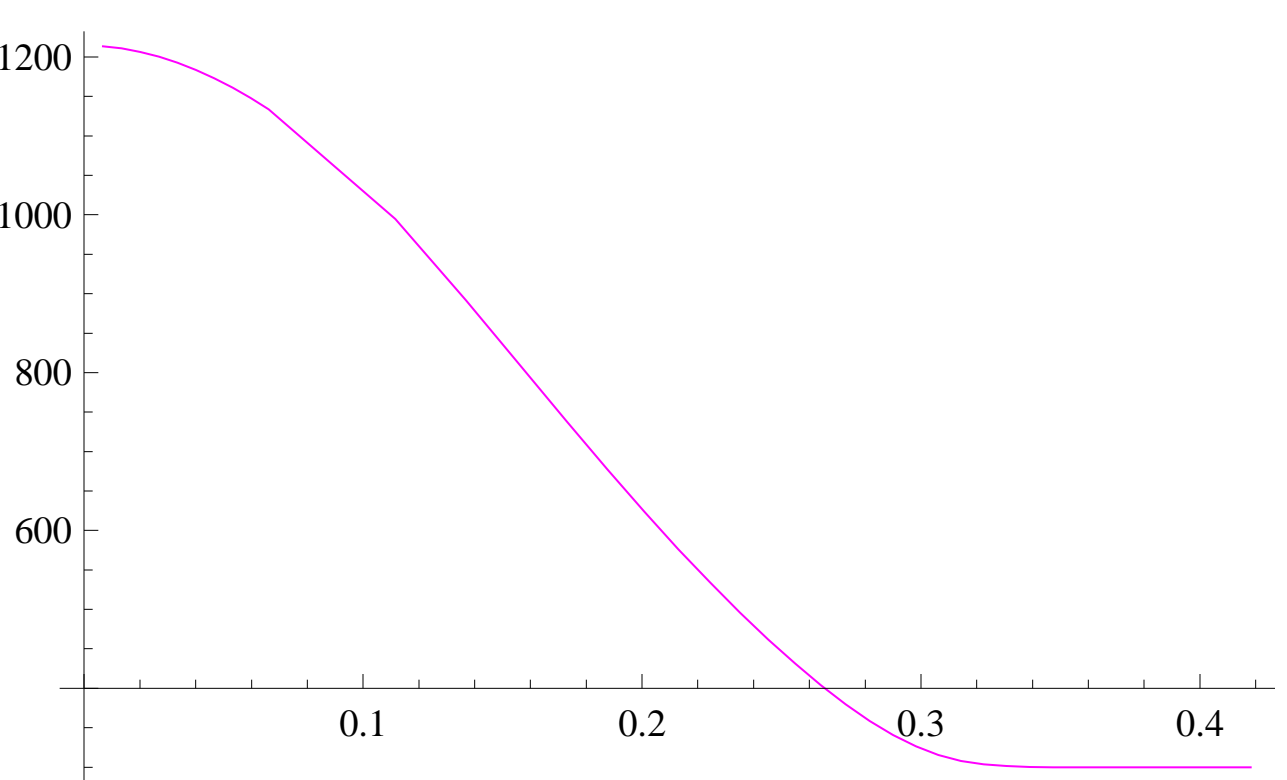

```
Show[%, %62, %60, %59, %58, %56, PlotRange → All]
```

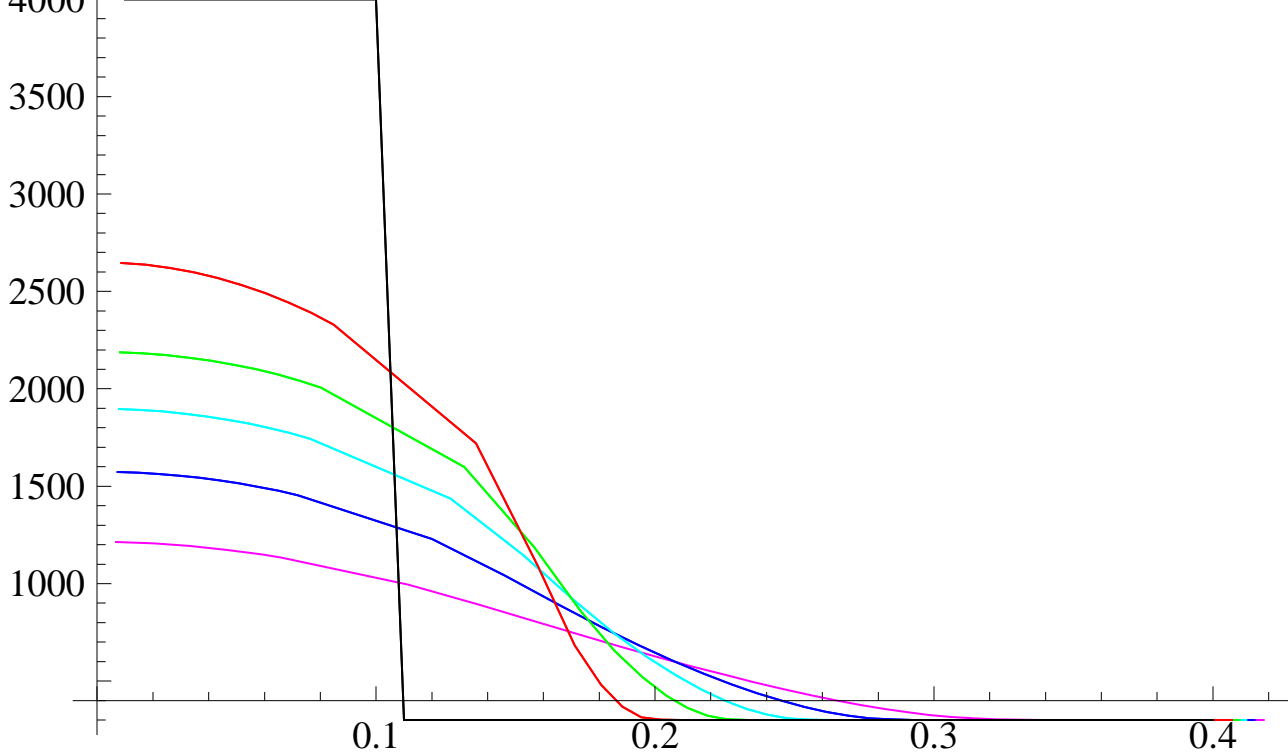

**Here we look at a large ball :**

```
Q = .1
```

```
0.1
```

```
Q
```

```
0.1
```

```
K = Table[Which[i ≤ 10, {.1 i, 4000}, 10 < i ≤ 40, {.1 i, 300}, True, {0, 0}], {i, 1, 41}]
```

```
{{0.1, 4000}, {0.2, 4000}, {0.3, 4000}, {0.4, 4000}, {0.5, 4000}, {0.6, 4000},
 {0.7, 4000}, {0.8, 4000}, {0.9, 4000}, {1., 4000}, {1.1, 300}, {1.2, 300}, {1.3, 300},
 {1.4, 300}, {1.5, 300}, {1.6, 300}, {1.7, 300}, {1.8, 300}, {1.9, 300}, {2., 300},
 {2.1, 300}, {2.2, 300}, {2.3, 300}, {2.4, 300}, {2.5, 300}, {2.6, 300}, {2.7, 300},
 {2.8, 300}, {2.9, 300}, {3., 300}, {3.1, 300}, {3.2, 300}, {3.3, 300}, {3.4, 300},
 {3.5, 300}, {3.6, 300}, {3.7, 300}, {3.8, 300}, {3.9, 300}, {4., 300}, {0, 0}}
```

```
NestList[Evolve, K, 90 001];
```

```
Table[CT[%24[[i]]], {i, 1, 90 001}];
```

```
ListPlot[%, PlotRange → All]
```

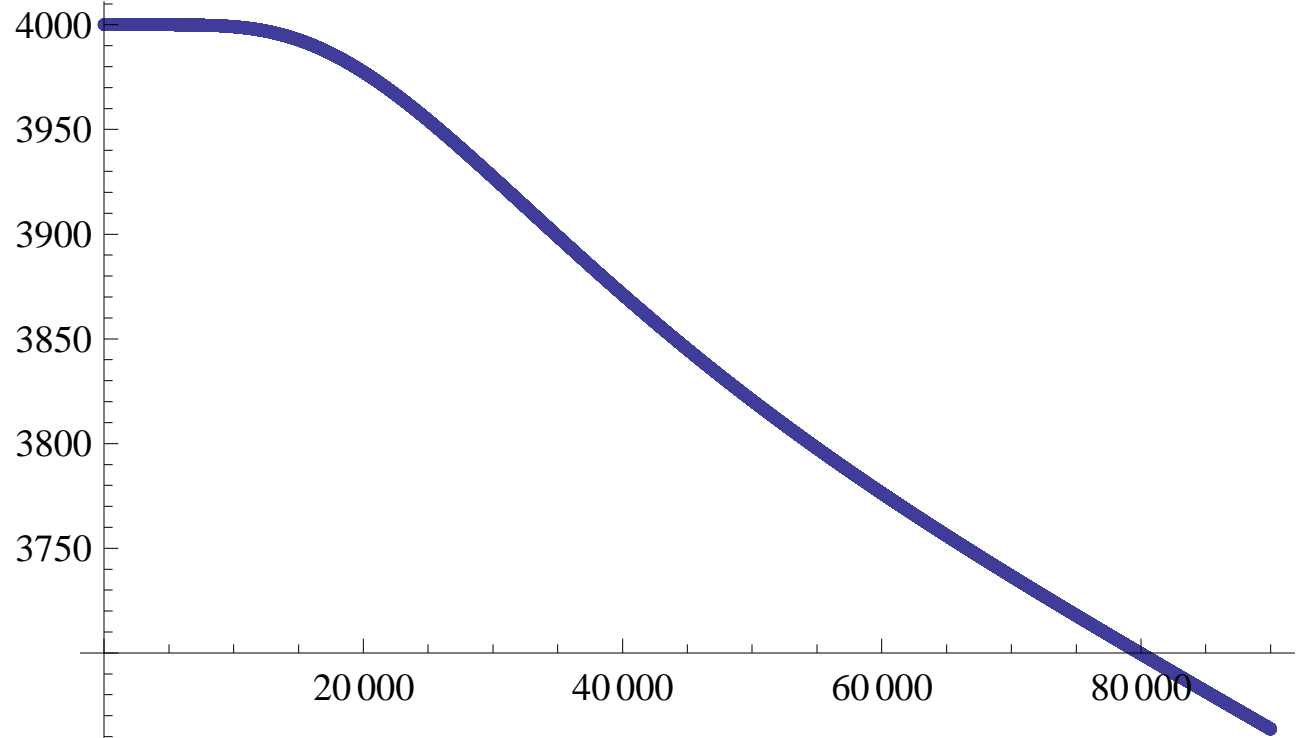

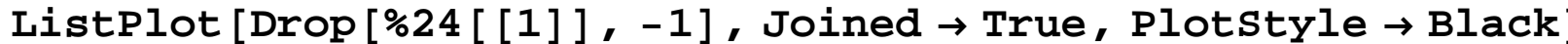

```
ListPlot[Drop[%24[[1]], -1], Joined → True, PlotStyle → Black]
```

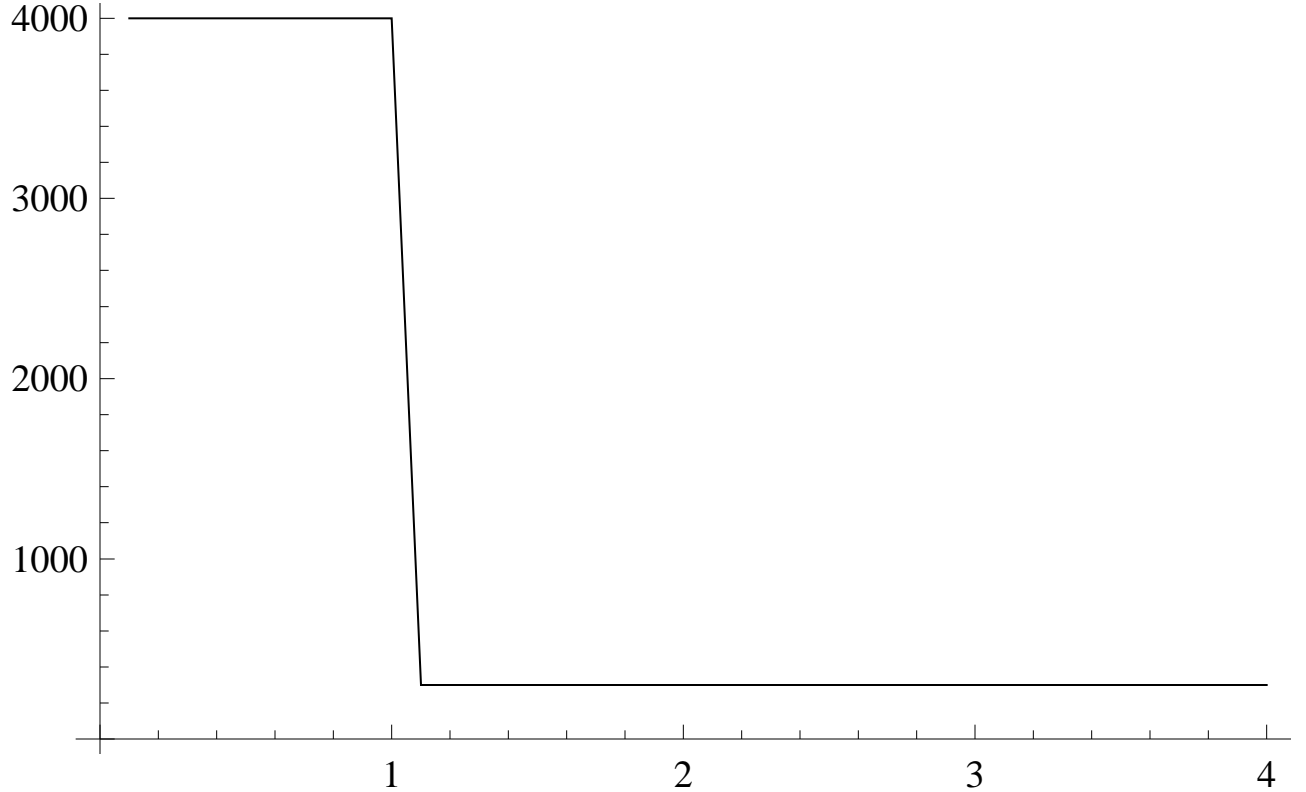

```
ListPlot[Drop[%24[[10 001]], -1], Joined → True, PlotStyle → RGBColor[1, 0, 0]]
```

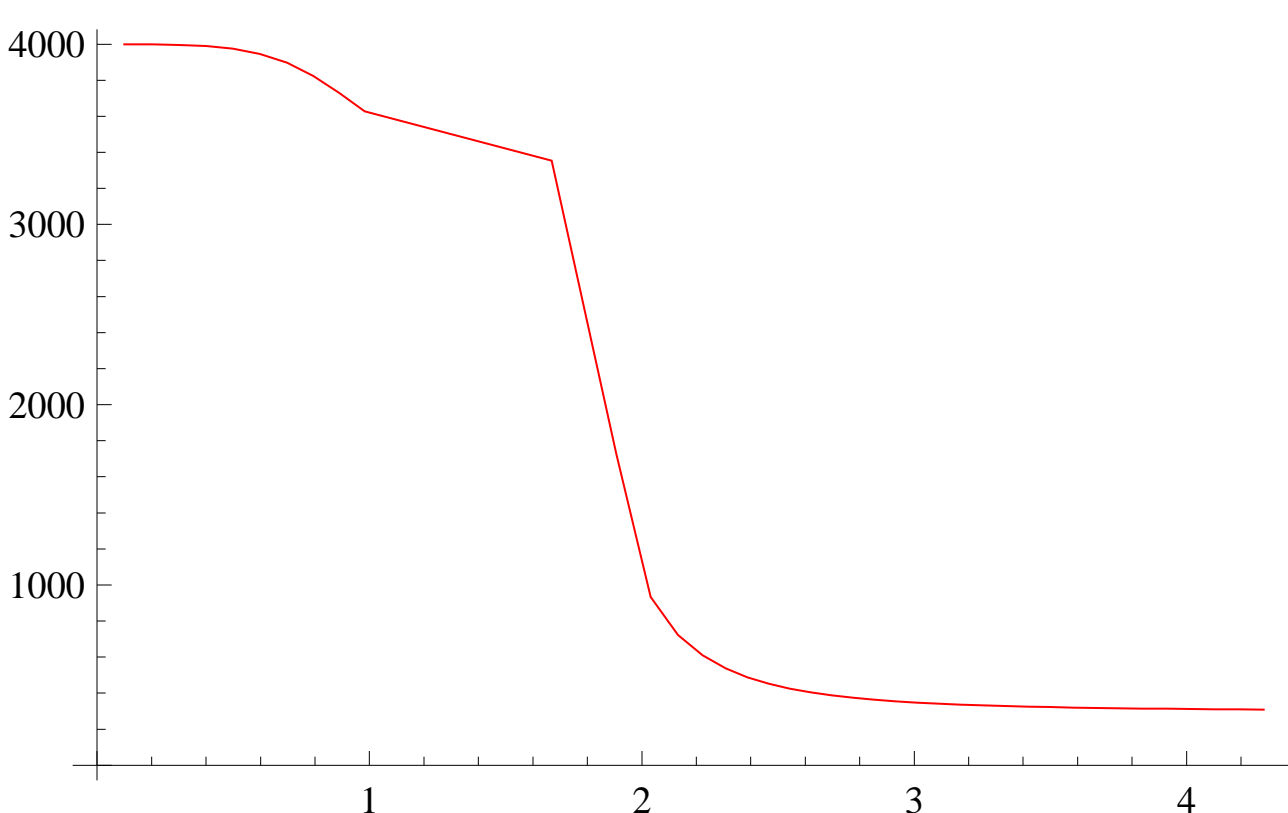

```
ListPlot[Drop[%24[[20 001]], -1], Joined → True, PlotStyle → RGBColor[0, 1, 0]]
```

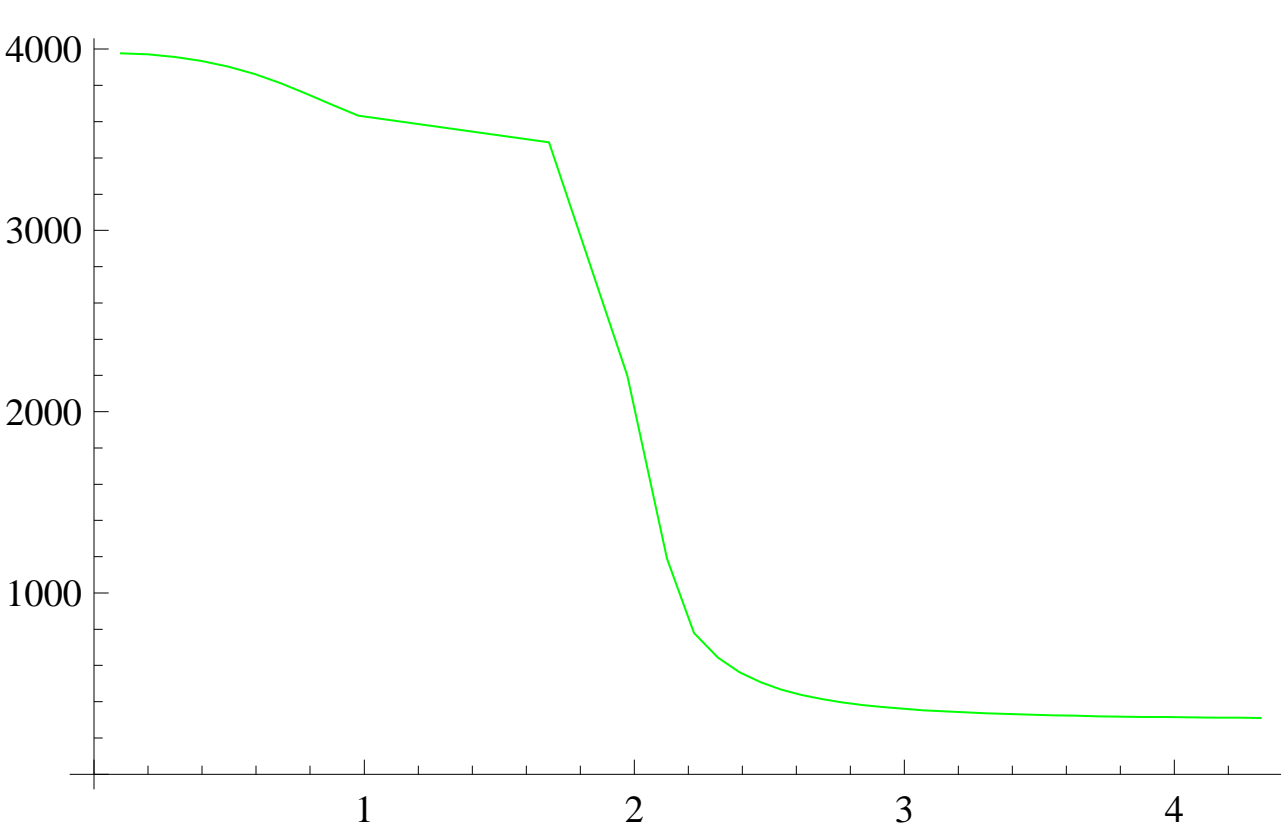

```
ListPlot[Drop[%24[[40 001]], -1], Joined → True, PlotStyle → RGBColor[0, 1, 1]]
```

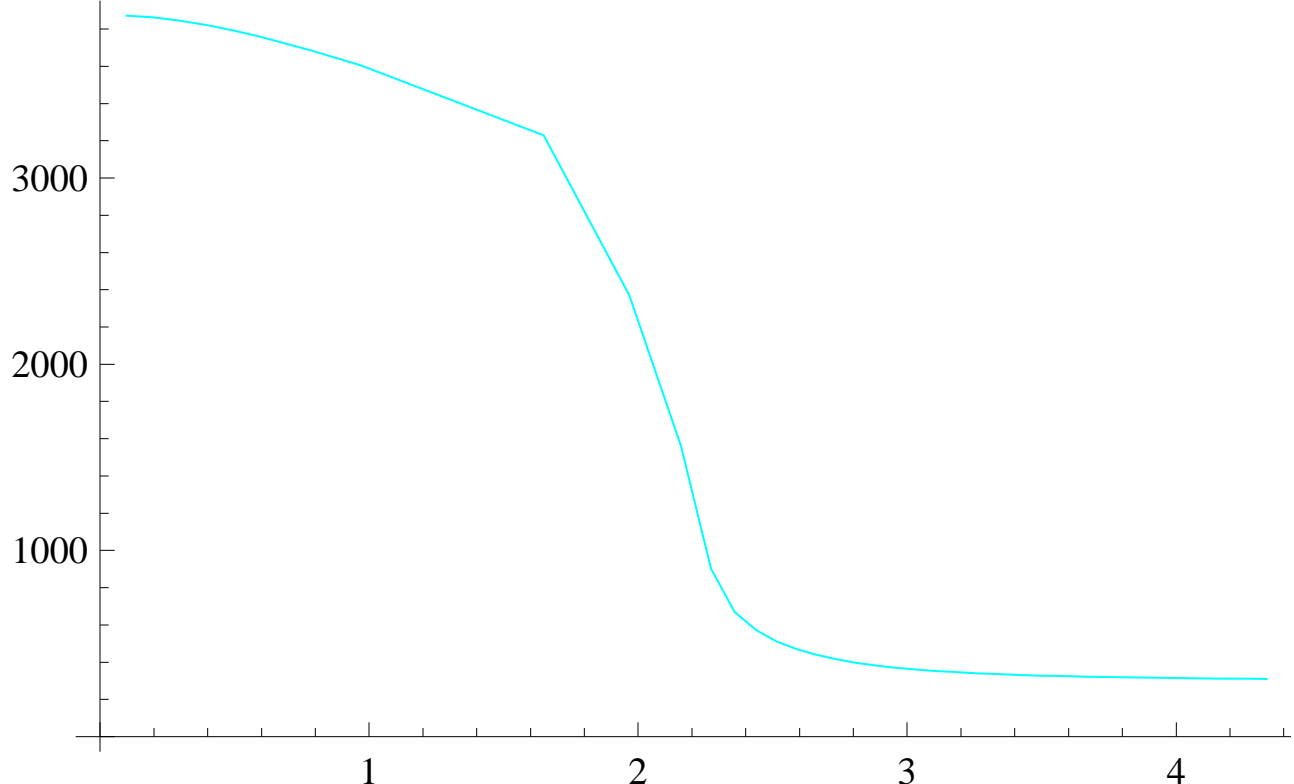

```
ListPlot[Drop[%24[[60 001]], -1], Joined → True, PlotStyle → RGBColor[0, 0, 1]]
```

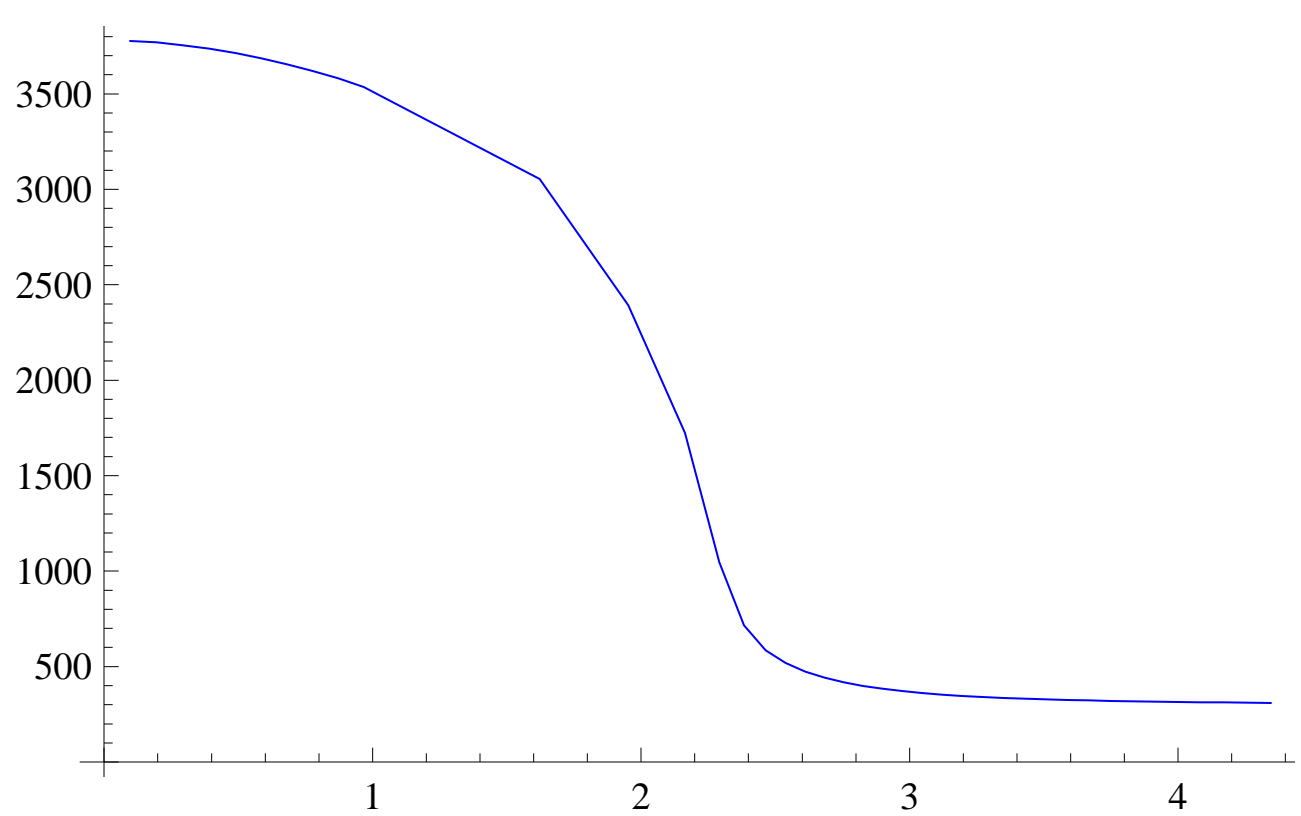

```
ListPlot[Drop[%24[[90 001]], -1], Joined → True, PlotStyle → RGBColor[1, 0, 1]]
```

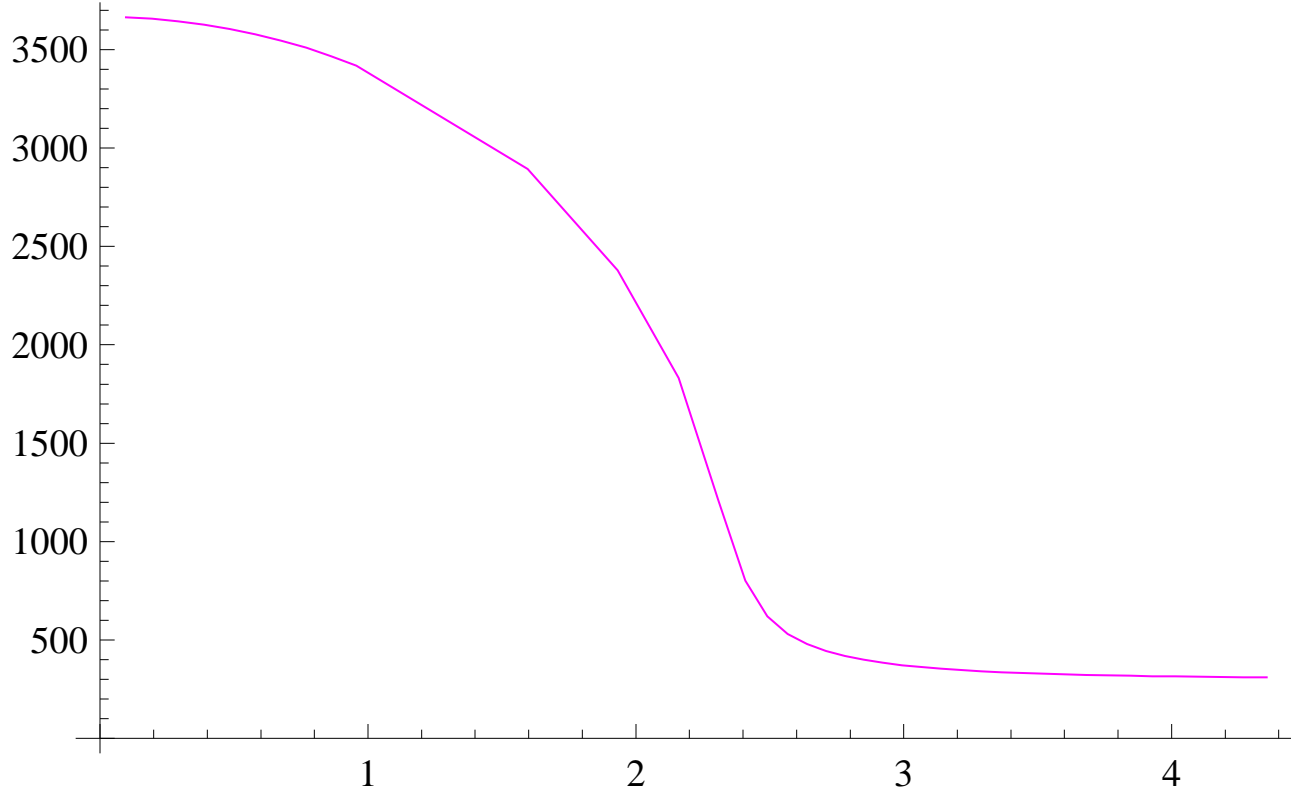

```
Show[%, %%, %%%, %%%%, %%%%%, %%%%%%, PlotRange → All]
```

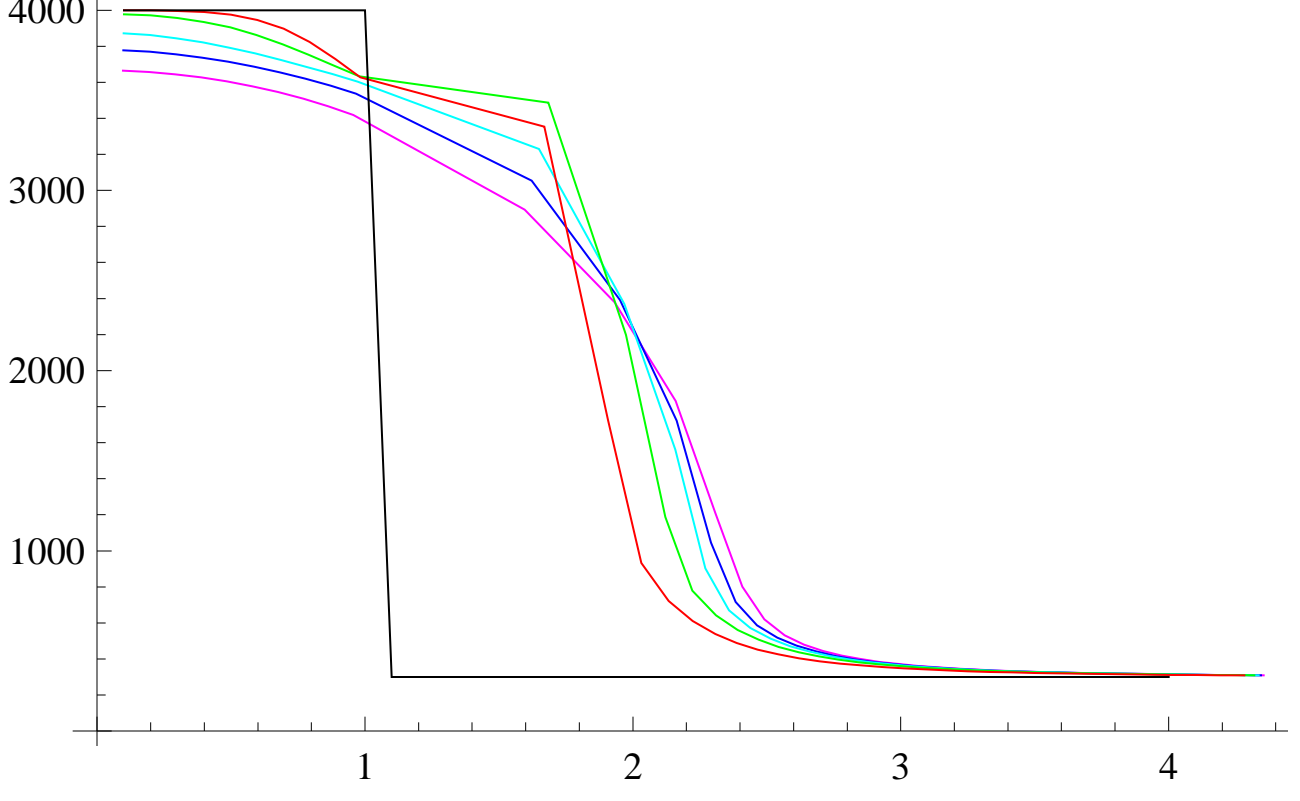

```
Lum[K]
```

```
375.546
```

**Here we do the small ball :**

```
Diff[U_] := Table[If[i ≤ 40,
   {U[[i]][[1]], U[[i]][[2]] + .0001 D0[U[[i]][[2]]] HH[U][[i]][[2]] + .0001 H[U][[i]][[2]]},
   U[[i]] + .0001 {0, 1}], {i, 1, 41}]
```

```
K = Table[Which[i ≤ 10, {.001 i, 4000}, 10 < i ≤ 40, {.001 i, 300}, True, {0, 0}], {i, 1, 41}]
```

```
{{0.001, 4000}, {0.002, 4000}, {0.003, 4000}, {0.004, 4000}, {0.005, 4000},
 {0.006, 4000}, {0.007, 4000}, {0.008, 4000}, {0.009, 4000}, {0.01, 4000}, {0.011, 300},
 {0.012, 300}, {0.013, 300}, {0.014, 300}, {0.015, 300}, {0.016, 300}, {0.017, 300},
 {0.018, 300}, {0.019, 300}, {0.02, 300}, {0.021, 300}, {0.022, 300}, {0.023, 300},
 {0.024, 300}, {0.025, 300}, {0.026, 300}, {0.027, 300}, {0.028, 300}, {0.029, 300},
 {0.03, 300}, {0.031, 300}, {0.032, 300}, {0.033, 300}, {0.034, 300}, {0.035, 300},
 {0.036, 300}, {0.037, 300}, {0.038, 300}, {0.039, 300}, {0.04, 300}, {0, 0}}
```

```
Source[U_] :=
 Table[Which[i ≤ 10, U[[i]], 10 < i < 41, .0001 {0, (706 Q^2 e^(-0.2` U[[41]][[2]]))/(β[U[[i]][[2]]] U[[i]][[1]]^4) U[[i]][[2]]} +
    U[[i]] - {0, 1} .0001 U[[i]][[2]] W[U[[i]][[2]]] / (P0 β[U[[i]][[2]]]),
   True, {0, 0} + U[[i]]], {i, 1, 41}]
```

```
Q = .00006
```

```
0.00006
```

```
NestList[Evolve, K, 100 001];
```

```
ListPlot[Drop[%25[[1]], -1], Joined → True, PlotStyle → Black]
```

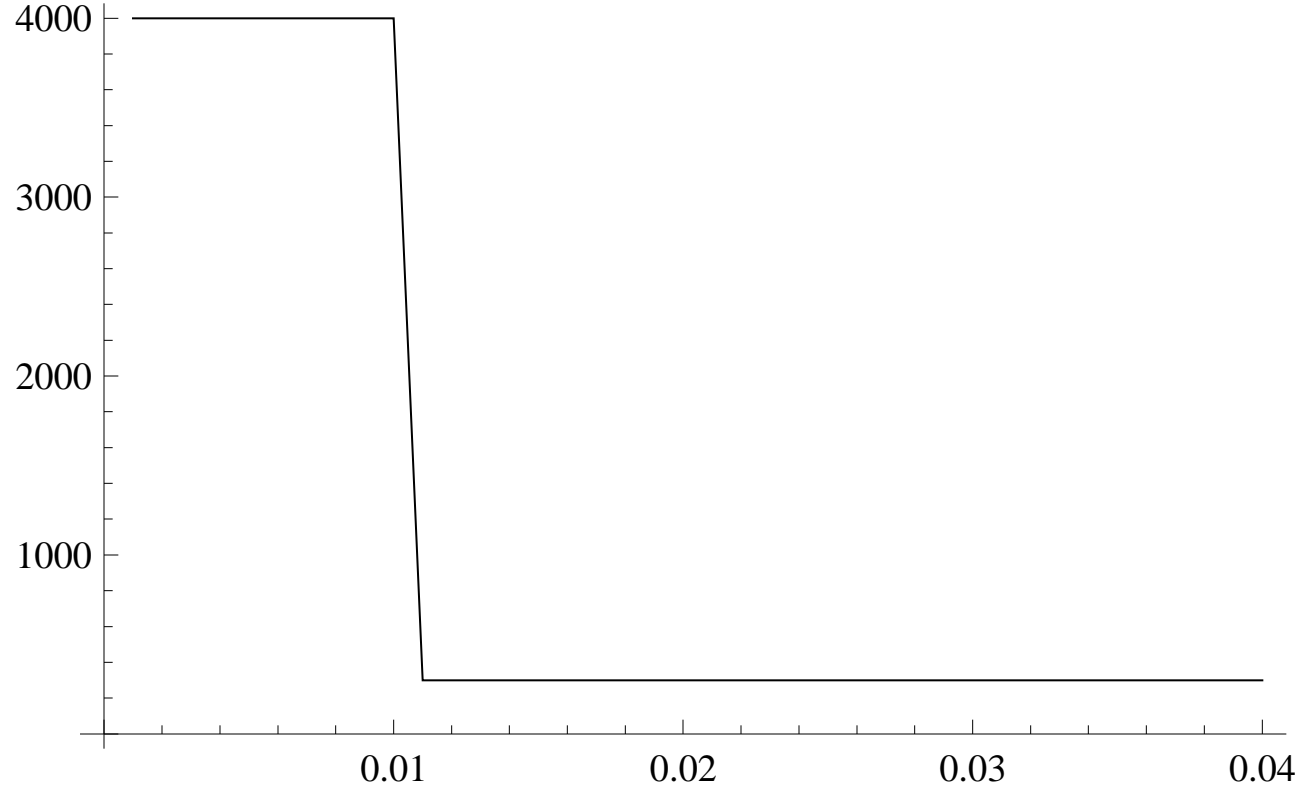

```
ListPlot[Drop[%25[[10 001]], -1], Joined → True, PlotStyle → RGBColor[1, 0, 0]]
```

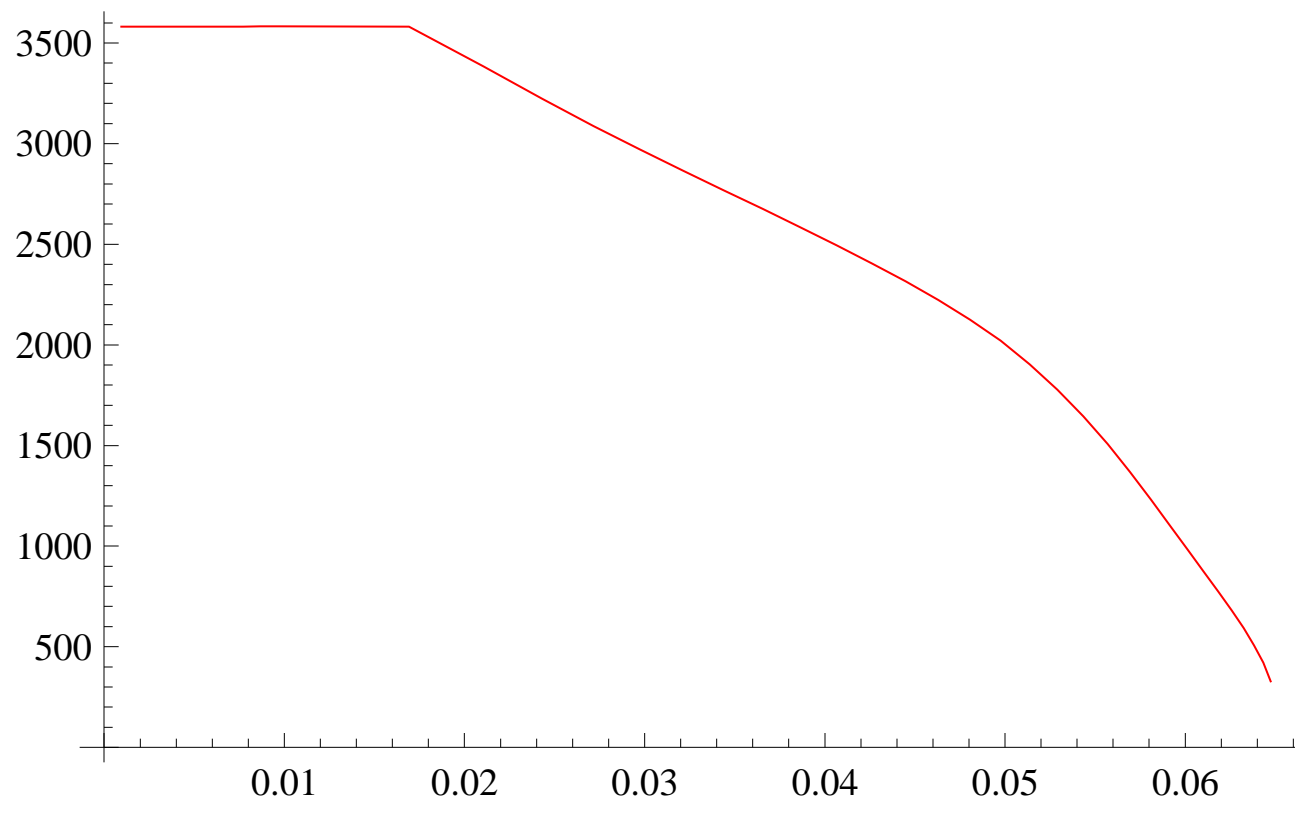

```
ListPlot[Drop[%25[[20 001]], -1], Joined → True, PlotStyle → RGBColor[0, 1, 0]]
```

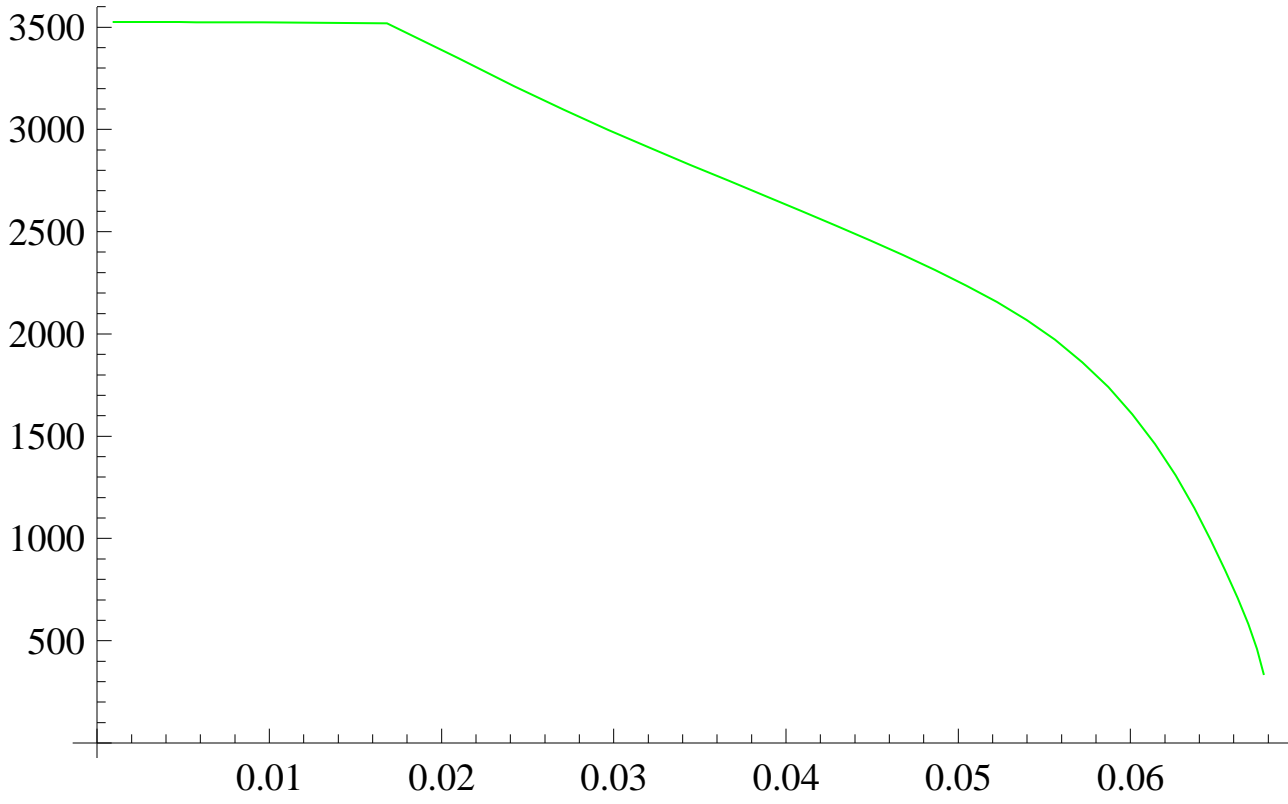

```
ListPlot[Drop[%25[[30 001]], -1], Joined → True, PlotStyle → RGBColor[0, 1, 1]]
```

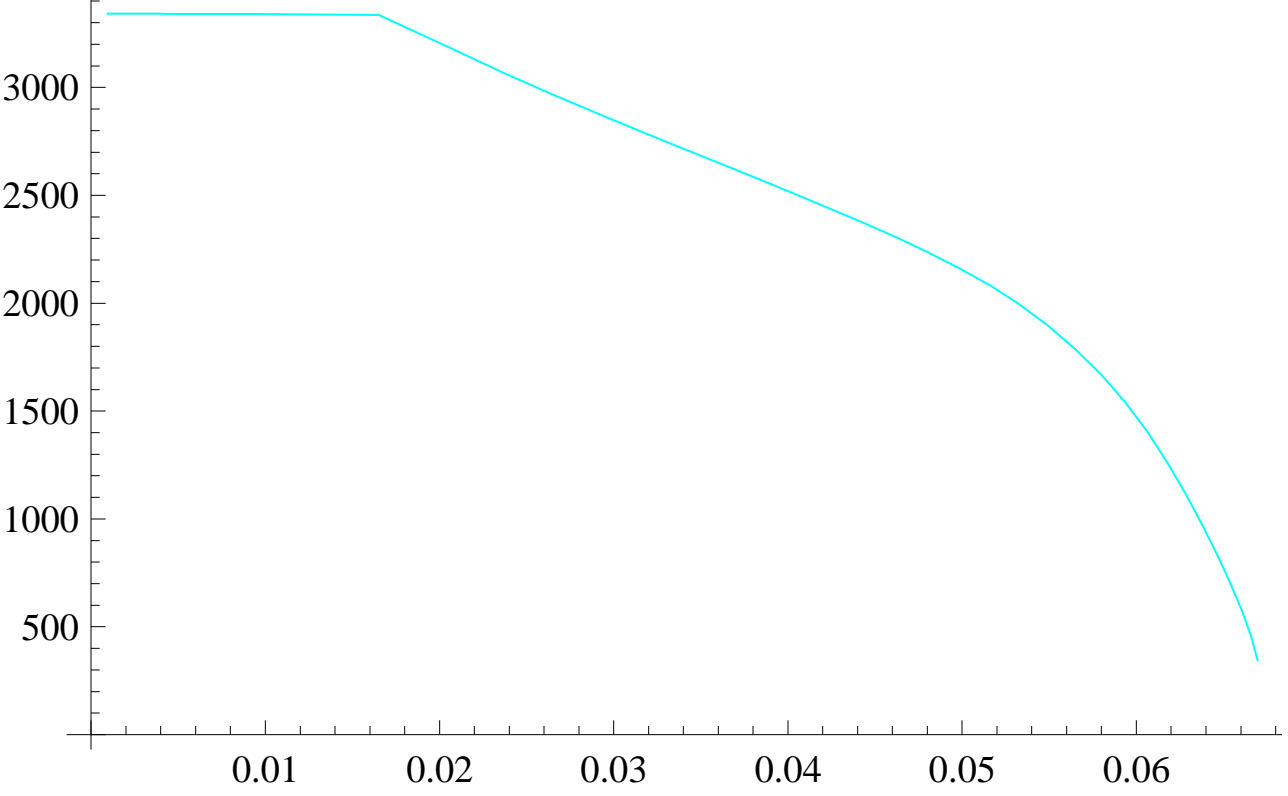

```
ListPlot[Drop[%25[[50 001]], -1], Joined → True, PlotStyle → RGBColor[0, 0, 1]]
```

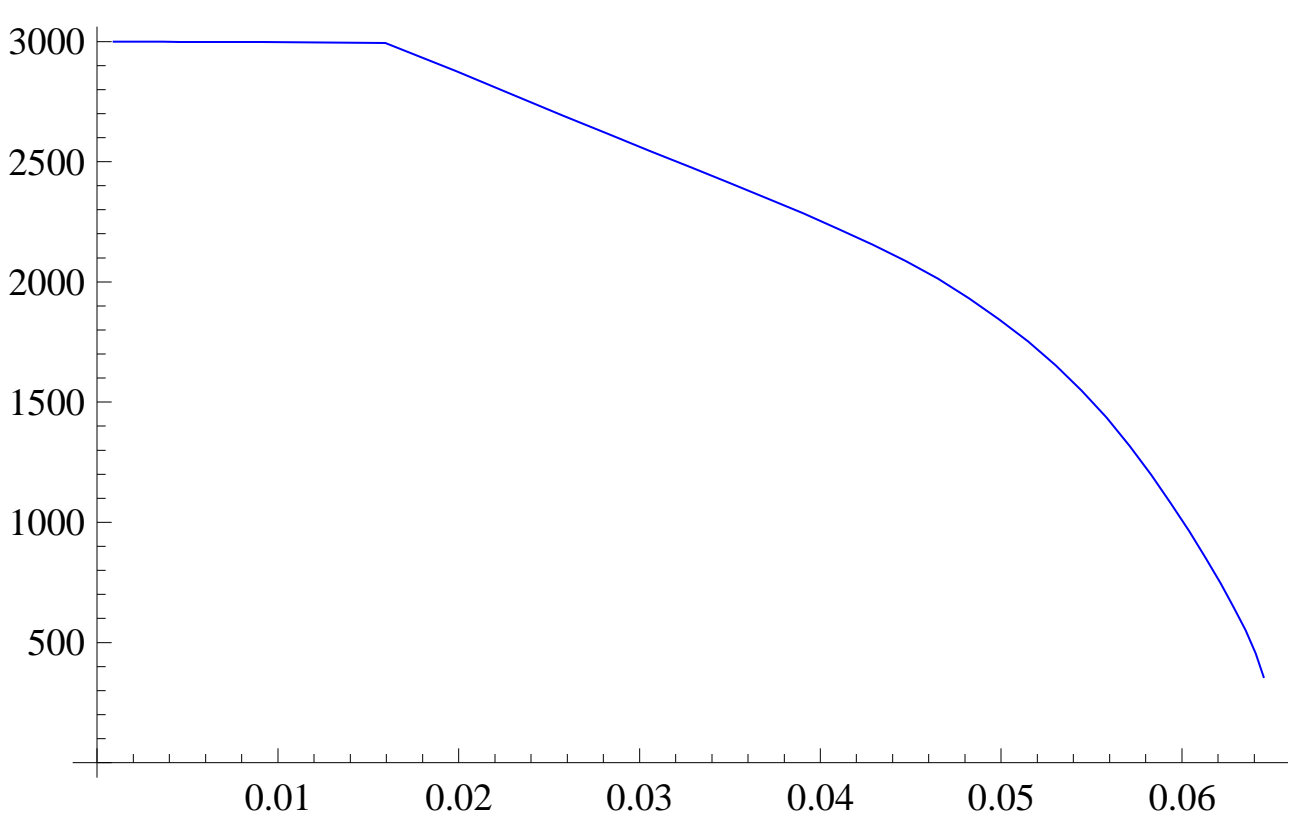

```
ListPlot[Drop[%25[[100 001]], -1], Joined → True, PlotStyle → RGBColor[1, 0, 1]]
```

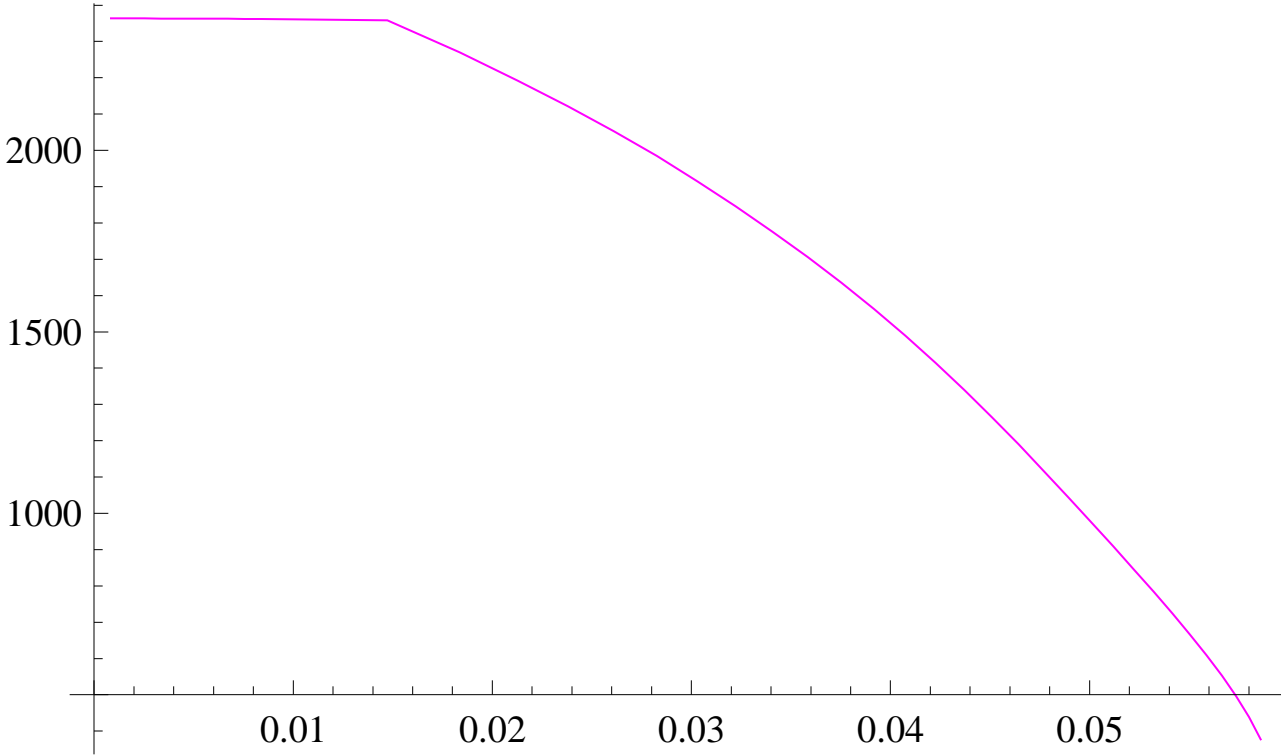

```
Show[%, %%, %%%, %%%%, %%%%%, %%%%%%, PlotRange → All]
```

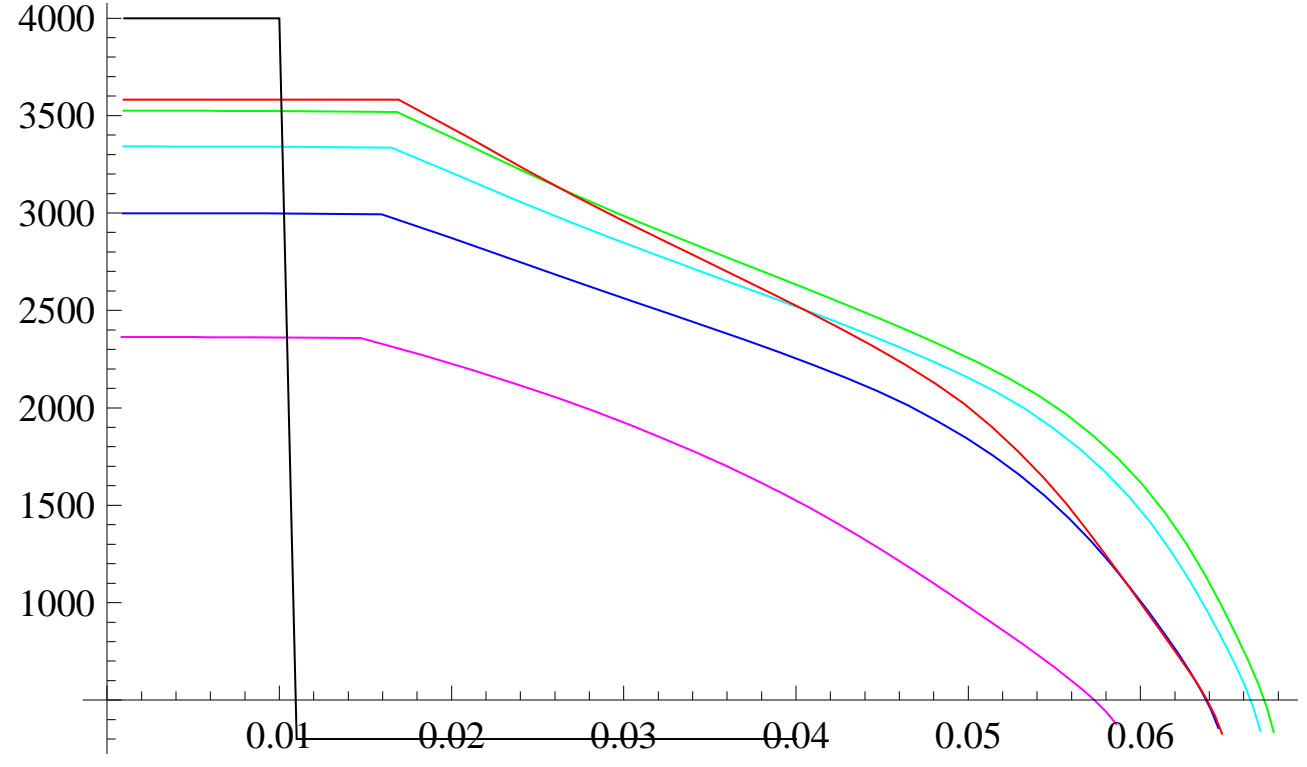

```
Lum[%25[[1]]]
```

```
0.000375546
```

```
Lum[%25[[10 001]]]
```

```
0.00288367
```

```
Lum[%25[[20 001]]]
```

```
0.0031522
```

```
Lum[%25[[30 001]]]
```

```
0.00209555
```

```
Lum[%25[[50 001]]]
```

```
0.000818929
```

```
Lum[%25[[100 001]]]
```

```
0.0000607975
```